\documentclass[12pt]{article}

\usepackage{amsmath}
\usepackage{amssymb}
\usepackage{amsthm}
\usepackage{graphicx}
\usepackage{natbib}
\setcitestyle{round}
\usepackage[hidelinks]{hyperref}
\usepackage{geometry}
\usepackage{setspace}
\usepackage{booktabs}
\usepackage{float}
\usepackage{xspace}
\xspaceaddexceptions{\%}
\usepackage{xurl}

\newcommand{\ArrCROneFromSzero}{49.0\xspace}
\newcommand{\ArrCRThreeFromSzero}{50.0\xspace}
\newcommand{\ArrCdrDelayYears}{1.0\xspace}
\newcommand{\BrownFusionPremiumDeployment}{-19.8\xspace}
\newcommand{\BrownFusionPremiumToday}{-19.4\xspace}
\newcommand{\BrownTerminationPremiumDeployment}{+0.5\xspace}
\newcommand{\CalAcidFloor}{20\xspace}
\newcommand{\CalActivation}{512\xspace}
\newcommand{\CalActivationGap}{85\xspace}
\newcommand{\CalBeta}{2.5\xspace}
\newcommand{\CalEtaMax}{25\xspace}
\newcommand{\CalEtaMaxYears}{4\xspace}
\newcommand{\CalGammaTargetEightHundred}{12.5\xspace}
\newcommand{\CalGammaZero}{0.0498\xspace}
\newcommand{\CalGridMax}{2000\xspace}
\newcommand{\CalKappaC}{0.05677\xspace}
\newcommand{\CalKappaF}{0.05207\xspace}
\newcommand{\CalLambdaBar}{0.5\xspace}
\newcommand{\CalLifetime}{200\xspace}
\newcommand{\CalOmegaTwo}{0.00439\xspace}
\newcommand{\CalPsiAtSzero}{1.4\xspace}
\newcommand{\CalRbar}{15\xspace}
\newcommand{\CalRbarPpm}{1.92\xspace}
\newcommand{\CalReferenceShutdown}{320\xspace}
\newcommand{\CalRiskAversion}{8\xspace}
\newcommand{\CalSigmaS}{1.0\xspace}
\newcommand{\CalSzero}{427\xspace}
\newcommand{\CalTcs}{2.3\xspace}
\newcommand{\CalTstar}{2.0\xspace}
\newcommand{\CdrPrizeRevival}{177.5\xspace}
\newcommand{\CurrentBrownGain}{6.5\xspace}
\newcommand{\CurrentTaxReduction}{3.4\xspace}
\newcommand{\FusionPrizeRevival}{40.8\xspace}
\newcommand{\GreenRevivalReversal}{940\xspace}
\newcommand{\GreenTermTRPCross}{650\xspace}
\newcommand{\InnSeta}{706\xspace}
\newcommand{\InnShat}{665\xspace}
\newcommand{\InnSstar}{667\xspace}
\newcommand{\LandingBothAtSixFiveZero}{64.0\xspace}
\newcommand{\LandingCdrAtSixFiveZero}{54.0\xspace}
\newcommand{\LandingNeitherAtSixFiveZero}{47.0\xspace}
\newcommand{\SCCLiveROneAtEightZeroZero}{149.4\xspace}
\newcommand{\SCCLiveROneAtFourTwoSeven}{70.4\xspace}
\newcommand{\SCCLiveROneAtSixZeroZero}{121.3\xspace}
\newcommand{\SCCLiveRTwoAtSixZeroZero}{32.9\xspace}
\newcommand{\SevDebtAtEightZeroZero}{1.48\xspace}
\newcommand{\SevLossAtSevenZeroZero}{5.35\xspace}
\newcommand{\SevLossAtSixZeroZero}{1.01\xspace}
\newcommand{\StallDownward}{644\xspace}
\newcommand{\WelBridgePeakStock}{686\xspace}
\newcommand{\WelDeterrencePositiveAbove}{698\xspace}
\newcommand{\WelNetNegativeAbove}{831\xspace}

\newtheorem{assumption}{Assumption}
\newtheorem{proposition}{Proposition}

\newtheorem{corollary}{Corollary}

\defcitealias{NASEM2019}{NASEM (2019)}

\title{\textbf{The Geoengineering Put: Ecological Debt, Termination Risk, and Asset Prices}}
\author{
    Samuel Hartzmark\thanks{Boston College, Carroll School of Management and NBER.} \\
    \and
    Kelly Shue\thanks{Yale University, School of Management and NBER.}
}
\date{November 16, 2026}

\begin{document}

\maketitle

\begin{center}
\end{center}

\begin{abstract}

Solar geoengineering offers an inexpensive and fast-acting way to lower global temperatures without removing carbon from the atmosphere. As warming intensifies, climate-vulnerable countries face growing incentives to deploy geoengineering unilaterally. The economic and financial consequences of this prospect remain largely overlooked. We develop a general equilibrium macro-finance framework in which geoengineering acts as a put option on climate damages, funded by the ``ecological debt'' of masked warming. If deployment stops, temperatures snap back in a catastrophic termination shock. Anticipation of geoengineering raises current brown firm valuations and lowers the optimal carbon tax, helping explain two puzzles in the existing literature: high brown-firm valuations and low carbon taxes. As ecological debt grows, however, termination risk transforms brown firms into toxic assets. Geoengineering also reshapes green innovation: it buys time for a breakthrough, but can weaken innovation incentives by lowering energy and carbon prices. Termination risk partly restores research incentives and directs them toward carbon removal, which can retire ecological debt. Our calibrations show that the geoengineering put can raise welfare across a range of parameter assumptions. Our framework clarifies how this conclusion depends on uncertain parameters and where further research is most needed.
\end{abstract}

\newpage

\section{Introduction} \label{sec:intro}

The economic framework to understand global warming typically revolves around the gargantuan costs to humanity of climate change (around \$10 trillion per year)\footnote{For example, \cite{bilal2026} estimate an impact of about 20\% of GDP, around \$20 trillion in today's dollars.} and the immense cost required to address the problem (around \$1 trillion per year).\footnote{For example, the IPCC estimates costs of about 1\% to 4\% of GDP, roughly \$1 to \$4 trillion \url{https://www.ipcc.ch/report/ar6/wg3/chapter/chapter-3/} } Further complicating the issue is the free-rider problem whereby sovereign actors face an incentive to under-invest in the transition to a low-carbon economy, as the benefits of one country's costly mitigation are globally shared. The potential deployment of climate geoengineering fundamentally alters this paradigm.

Geoengineering offers a physically plausible way to cool the planet that is orders of magnitude less expensive than commonly discussed remedies to climate change. While geoengineering decreases the earth's temperature, it does not reduce the underlying atmospheric carbon stock. By simply masking the symptoms of accumulating carbon, geoengineering introduces the risk of a termination shock: cessation of geoengineering would induce a rapid spike in global temperatures dictated by the high underlying carbon stock, with potentially massive economic costs. 

Geoengineering's low cost and ease of implementation make it feasible for a single country to deploy unilaterally. This shifts the climate dilemma from a free-rider problem to what \citet{Weitzman2015} termed the free-driver problem, where one actor can lower the global thermostat. Countries facing the most severe climate damages have the strongest incentives to act. A government facing millions of deaths from extreme heat or flooding may not be deterred by sanctions or other penalties. The credible threat of unilateral deployment may also induce other countries to coordinate deployment before the vulnerable country acts alone. Whether deployment occurs unilaterally or through such cooperation, the growing incentive for climate-vulnerable countries to act makes geoengineering increasingly likely as global warming intensifies.

The implications of geoengineering risk have been largely overlooked in the climate finance literature, which has instead focused on transition risk. Transition risk refers to the repricing of assets that can occur as policy or technological change moves the economy away from fossil fuels. Aggressive carbon pricing and environmental regulation can erode the cash flows of carbon-intensive firms and impair the fiscal capacity of sovereigns whose tax bases depend on fossil fuel extraction. A growing literature explores the asset pricing implications of transition risk across equities, corporate debt, and sovereign markets \citep{BoltonKacperczyk2021, BoltonKacperczyk2023, vanderPloegRezai2020, EngleGiglioKellyLeeStroebel2020, AcharyaGiglioPastoreStroebelTanYong2025, BarnettBrockHansen2020}. A more recent strand examines biodiversity risk, the financial exposure of firms and sovereigns to the loss of natural capital and the ecosystem services on which production depends \citep{Dasgupta2021, GiglioKuchlerStroebelWang2025, GarelRomecSautnerWagner2024}. Geoengineering constitutes another important source of climate-related risk, with asset pricing implications that are sharply different from those of transition risk.

In this paper, we develop a macro-finance framework to study the economic implications of climate geoengineering in a general equilibrium setting. We define geoengineering broadly as any technological intervention that lowers realized global temperatures without altering the underlying atmospheric carbon stock. Our calibration estimates focus on solar radiation management (SRM), a class of technologies that cool the planet by reflecting a small fraction of incoming sunlight. The most prominent example is stratospheric aerosol injection (SAI), which disperses reflective particles such as sulfur dioxide into the stratosphere.\footnote{Our definition of geoengineering also applies to other SRM approaches including marine cloud brightening, surface or cloud albedo enhancement, and space-based reflectors, as well as radiative forcing technologies, such as cirrus cloud thinning.} The physics of SAI are illustrated by the eruption of Mount Pinatubo in 1991 that injected about 20 million tons of sulfur dioxide into the stratosphere, cooling the planet by about 0.5 degrees Celsius over the next two years. Engineering estimates put direct delivery costs at a few billion dollars per year for an early program \citep{SmithWagner2018, McClellanKeithApt2012}, and \$10 to \$20 billion per year at full scale \citep{smith2020cost}--- a small fraction of the annual capital expenditure required for global decarbonization. Given this asymmetric cost structure, a sovereign nation facing severe heat damages may possess both the means and the incentive to act unilaterally.

We formalize the macroeconomic consequences of geoengineering deployment by introducing two organizing concepts. The first is the ``geoengineering put.'' Deployment caps global temperatures and truncates the right tail of climate damages, acting as a put option on the climate. However, because geoengineering only masks the warming effects of greenhouse gases without removing them from the atmosphere, exercise of this put is accompanied by the accumulation of ``ecological debt'': the gap between the realized temperature and the temperature the world would experience given the carbon stock in the absence of geoengineering. Thus, deployment carries an accumulating catastrophic tail risk.

This tail risk is recognized in the geophysical literature as a termination shock \citep{ParkerIrvine2018, TrisosCriscitielloNicholls2018}. A termination shock occurs because of the difference in lifespans between carbon dioxide and stratospheric aerosols. Carbon dioxide lingers in the atmosphere for centuries whereas stratospheric aerosols dissipate in one to two years. If geoengineering were to be suddenly halted, temperatures would snap back to the level determined by the underlying carbon stock, and the speed of the warming would greatly exceed that of baseline climate change. This sudden jump in global temperatures could lead to catastrophic destruction of global capital and productivity. In our framework, a termination shock represents a default on ecological debt.

We organize our analysis into three nested regimes, which successively activate the main economic mechanisms. In the baseline (Regime 1), geoengineering does not exist and the economy faces standard climate risks. In Regime 2, geoengineering is deployed once the carbon stock passes a known threshold. Continuation of geoengineering is certain, allowing us to isolate the moral hazard channel. In the full model (Regime 3), we introduce a termination hazard. Our key insight is that the same physical model of geoengineering can have opposite effects on asset prices at high carbon stocks depending on whether termination risk is present.

We numerically solve the planner's Hamilton-Jacobi-Bellman (HJB) equation to obtain the forward-looking social cost of carbon, equilibrium dynamics, and asset prices for each regime. The model generates four main insights regarding how termination risk affects climate policy and asset prices and a second set of results regarding how geoengineering and the termination risk it creates affect innovation.

First, the geoengineering put acts as a moral hazard device. It reduces realized temperatures, which decreases the optimal carbon tax, raises emissions and the value of fossil fuels, and slows the green transition. If geoengineering were to mask all warming perfectly and permanently, the price of temperature damages would disappear from the social cost of carbon, and climate risk would no longer be priced through temperature damages.

Second, termination risk fundamentally alters the cross-section of expected returns. Firms that lose value when termination lowers consumption carry a positive termination risk premium, while firms that gain value can provide insurance. We show that the value of fossil fuel firms is non-monotonic in the carbon stock. At first, geoengineering helps fossil fuels by lowering the carbon tax. But eventually, it makes them toxic assets as ecological debt builds up. The potential for a clean energy breakthrough (e.g., fusion) introduces a different risk: competition from the new energy supplier can lower both Brown and Green firm values.

Third, the relationship between geoengineering and carbon dioxide removal (CDR) flips. Without termination risk, geoengineering crowds out CDR research by suppressing the carbon prices that reward innovators. With termination risk, geoengineering and CDR can become complements because CDR can pay down ecological debt, reducing the damages from termination risk.

Fourth, the social cost of carbon can be higher under geoengineering than it would be without geoengineering. At high enough concentrations, the next unit of carbon becomes more socially costly in Regime 3 than in Regime 1, because it increases the damage if termination occurs.

The paper's next major contribution is to show that geoengineering and the termination risk it creates influence incentives for green innovation. One of the most compelling arguments in favor of solar geoengineering is peak shaving: deploy SRM to buy time while the world waits for a technological breakthrough \citep{KeithMacMartin2015}. We model two potential breakthroughs: cheap fusion energy (representing any new source of inexpensive, clean energy) and the emergence of inexpensive carbon removal at scale. The arrival rates have a background component and a component that depends on the market value of the innovation prize. Fusion also increases the probability of a CDR breakthrough by providing the cheap energy that some removal technologies may require. We show that, while geoengineering can serve as peak shaving, it can reduce incentives to innovate by suppressing energy and carbon prices.

Termination risk partly revives innovation incentives and redirects them. Without termination risk, geoengineering suppresses both green energy and carbon removal research. With termination risk, carbon removal research revives disproportionately, because removal is the only technology that can retire ecological debt. Revival can be late, however. Under our baseline calibration, when geoengineering is deployed, it holds down the amount of removal research (relative to no geoengineering) until the carbon stock rises to \InnSeta\ ppm.

A common hope is that geoengineering buys time for technological breakthroughs that allow a safe landing: after deployment, the world brings the carbon stock low enough to switch off geoengineering with nonpositive net emissions, before termination occurs. We show that even a CDR breakthrough does not ensure this outcome. The challenge is a stock-and-flow problem: removal must draw down accumulated carbon while keeping pace with new emissions, and annual removal capacity is limited. Moreover, CDR lowers the carbon price, encouraging new emissions that can offset removal and prolong dependence on geoengineering.

Finally, we compare welfare in a regime with geoengineering and termination risk to welfare in a regime without geoengineering. Our baseline calibration incorporates several pessimistic assumptions about termination risk. Although governments would have strong incentives to prevent interruptions as ecological debt grows and to restart geoengineering if halted, we impose a constant termination hazard of 0.5\% per year and rule out redeployment after termination. We also assume that termination permanently reduces productivity, with losses increasing in ecological debt and reaching 12.5\% at a carbon concentration of 800 ppm, in addition to the damages from renewed warming. Our assumptions about innovation incentives also lean pessimistic. Governments offer no additional prizes to encourage green energy or carbon removal breakthroughs. Instead, innovators’ rewards depend on market profits tied to energy and carbon prices, which can fall as innovation succeeds.

We find that welfare is higher with geoengineering under this baseline calibration and across a range of alternative parameter assumptions. Under the baseline, a breakthrough often arrives before deployment starts. When deployment does begin, the capped temperatures buy decades of avoided damages before ecological debt becomes large. The welfare ranking can reverse, however, when geoengineering has a very high termination hazard combined with substantial termination losses at moderate carbon stocks.

Our findings inform an active debate over proposals to restrict geoengineering research and deployment. To the extent that geoengineering has been addressed by governmental groups (e.g. the UN, Germany, the EU), the recommendations have been to prohibit experimentation and create international agreements prohibiting its implementation.\footnote{The UN Human Rights Council Advisory Committee has recommended prohibiting geoengineering development and restricting outdoor experiments \citep{UNHRCAdvisoryCommittee2023}, while the German Environment Agency recommends an international non-use agreement and a prohibition on field experiments in principle \citep{GermanEnvironmentAgency2025}. Mexico has likewise announced measures to prohibit experimentation \citep{SEMARNATConacyt2023}.} Our results challenge the presumption underlying these proposals that preventing geoengineering (by moving the world to Regime 1) would necessarily improve welfare.

We emphasize that our calibration and welfare analysis depend on parameters that remain poorly understood, including the timing of deployment, the arrival rate of breakthroughs, termination risk, and the response of innovation to carbon prices. These uncertainties reinforce the value of further research, including research into risks that could change a vulnerable country's decision to deploy.\footnote{Those who oppose geoengineering because they believe its consequences would be catastrophic also have a strong reason to support such research. A vulnerable country facing severe climate damages has its own incentive to deploy. To change that decision, research would need to show that the country faces greater harm from deployment than from continued warming.} Our primary contribution is a flexible framework for evaluating these competing effects. It identifies which uncertainties matter for welfare and allows the calibration to be updated as better parameter estimates become available.

Our paper contributes to several strands of the literature. First, we extend the asset pricing literature on climate-related risks \citep{GiglioKellyStroebel2021,AcharyaGiglioPastoreStroebelTanYong2025,GiglioKuchlerStroebelWang2025, BarnettBrockHansen2020} to a setting with geoengineering and termination risk. Following research on rare disasters \citep{Barro2006, Wachter2013, Gabaix2012}, our model features a stochastic discount factor with endogenous jump risk. The risk-neutral magnification of the termination hazard can generate large risk premia at high carbon stocks despite low physical probabilities of termination. In our setting, this magnification links the pricing of disaster risk to the accumulation of carbon under geoengineering.

Our paper also builds on the integrated assessment literature, which studies the social cost of carbon (SCC) and optimal carbon taxation \citep{Nordhaus1992,Nordhaus2017,GolosovHasslerKrusellTsyvinski2014,Barrage2020,LemoineTraeger2014,CaiLontzek2019}. A small but growing literature studies geoengineering using integrated assessment models and other theoretical frameworks. This research examines learning about effectiveness and damages \citep{MorenoCruzKeith2013,HeutelMorenoCruzShayegh2018,KellyEtAl2024}, climate tipping points and termination risk \citep{HeutelMorenoCruzShayegh2016,HelwegenEtAl2019}, and deployment and carbon taxation under strategic interaction and uncertainty \citep{MeierTraeger2022,MeierTraeger2023}. These studies also explore how geoengineering changes the optimal timing and scale of mitigation and carbon removal \citep{BelaiMorenoCruzKeith2024,DrakeEtAl2021} and the incentives to invest in clean technology \citep{AcemogluRafey2023}.

We build on these insights by incorporating geoengineering, termination risk, endogenous innovation, and the free driver problem into a quantitative general equilibrium framework that jointly determines carbon prices, financial valuations, and incentives for technological breakthroughs. This allows us to study how these mechanisms interact over time. Carbon prices and risk premia determine innovation rewards, innovation affects how long the world remains exposed to termination risk, and that risk feeds back into carbon prices and asset values as ecological debt accumulates. Financial valuations therefore both reflect climate risks and determine incentives to develop the technologies that reduce them. Our framework quantifies these interactions and evaluates their combined welfare implications, including how the carbon-price response that initially supports fossil fuel valuations can weaken innovation incentives and prolong dependence on geoengineering.

Our deployment assumptions complement earlier work by \citet{KeithMacMartin2015} and \citet{MacMartinCaldeiraKeith2014}, who develop a peak-shaving scenario with early, moderate deployment under cooperative governance. We also examine when a social planner would choose to deploy geoengineering and how much further warming it would offset. Under our baseline calibration, welfare is highest among the rules considered with immediate deployment that prevents all further warming. Yet the world is not deploying geoengineering today. This gap suggests that frictions outside our model may prevent deployment from following the social planner's preferred rule. We therefore focus on a potentially more realistic trigger: a vulnerable nation's credible threat of unilateral deployment.

Our paper underscores the importance of considering the implications of geoengineering today, even if deployment remains years away. Anticipation of geoengineering raises current fossil fuel valuations and lowers the optimal carbon tax, potentially helping explain two apparent puzzles: carbon taxes below the calibrated optimum in \citet{Nordhaus2019}, and brown-firm valuations that remain high despite substantial carbon burdens \citep{PastorStambaughTaylor2026,GreenstoneLeuzBreuer2023,AhnDuttaPatatoukas2025,Pedersen2026}. Under our baseline calibration, accounting for geoengineering increases current brown-firm value by \CurrentBrownGain\% and reduces the SCC by roughly \CurrentTaxReduction\% relative to a world without geoengineering. The reduction in the SCC is much larger under earlier deployment rules. More broadly, our analysis highlights the value of further research: understanding geoengineering's risks is valuable well before deployment, because expectations of those risks shape current asset values and the optimal response to climate change.

The remainder of the paper proceeds as follows. Section~\ref{sec:model} develops the model. Section~\ref{sec:results} presents the theoretical results. Section~\ref{sec:numerical} presents quantitative results from the solved model. Section~\ref{sec:conclusion} concludes.

\section{Model} \label{sec:model}

This section presents the model along with motivation for parameter estimates which will be used in the calibration. In Section~\ref{subsec:carbon_climate}, we first describe the physical climate system. Then, in Section~\ref{subsec:geo_debt}, we introduce geoengineering, ecological debt and termination risk. In Section~\ref{subsec:firms}, we derive firm dividends from energy production and prices. Section~\ref{subsec:planner} introduces the planner's problem in three steps. In Sections~\ref{subsec:breakthroughs}-\ref{subsec:prearrival_hjb}, we complete the planner's problem by adding the two breakthrough technologies, endogenous innovation and the post-arrival regime tree. In Section~\ref{subsec:asset_pricing}, we develop the asset pricing framework.

\subsection{Carbon Stock and Climate} \label{subsec:carbon_climate}

The main slow-moving state variable is the stock of carbon in the atmosphere, $S_t$, measured in parts per million (ppm):
\begin{equation} \label{eq:carbon_stock}
    dS_t=(B_t-R_t)\,dt+\sigma_S\,dZ_t.
\end{equation}
Here $B_t$ is emissions from Brown energy production and $R_t$ is carbon removal, both expressed in ppm per year. We set $\sigma_S=\CalSigmaS$ ppm/$\sqrt{\mathrm{yr}}$ to capture small carbon-cycle fluctuations, motivated by observed interannual variability \citep{Friedlingstein2023}.

Current novel carbon removal is negligible relative to annual emissions of roughly 40 GtCO$_2$ \citep{Friedlingstein2023}. We therefore round removal capacity before its breakthrough to zero.\footnote{We convert emissions and removal measured in GtCO\(_2\) into changes in atmospheric CO\(_2\) concentration using the same factor of approximately 7.81 GtCO\(_2\) per ppm. We also omit passive natural uptake rather than applying different conversion factors during accumulation and drawdown. This is a one-stock approximation to the carbon cycle--for a richer model that incorporates natural sinks and the persistence of atmospheric carbon, see \cite{Joos2013}.}

Absent geoengineering, global temperature anomaly is a strictly increasing function of the carbon stock:
\begin{equation} \label{eq:temp_unmitigated}
    T_t^{un} = \Psi(S_t), \qquad \Psi' > 0
\end{equation}
Throughout the paper, temperature refers to warming relative to the preindustrial level, so today's temperature corresponds to $T \approx \CalPsiAtSzero^\circ$C, and the 280 ppm appearing in $\Psi$ below is the preindustrial carbon stock. This increasing function captures the logarithmic relationship between CO$_2$ concentration and radiative forcing, combined with climate feedbacks that amplify warming.

We use a reduced-form concentration-temperature mapping, $\Psi(S)=\CalTcs\log_2(S/280)$, chosen so that $\Psi(\CalSzero)=\CalPsiAtSzero^\circ$C at the reference stock $S_0=\CalSzero$ ppm. This is a simplification of integrated assessment models. DICE-2023 includes gradual temperature adjustment and ocean heat uptake \citep{BarrageNordhaus2024}; our one-state mapping omits that history dependence, including different temperatures at the same stock during accumulation and drawdown.

We decompose climate damages to aggregate productivity into a temperature component and a carbon-stock component,
\begin{equation} \label{eq:damage}
    \Omega(T) + \Omega_{acid}(S), \qquad \Omega(T) = \omega(T) + \zeta \cdot h(T - \bar{T})
\end{equation}
where $\omega(T)$ represents standard climate damages (e.g., heat stress or agricultural yield losses) and is smooth, increasing, and convex. $\bar{T}$ is a temperature threshold associated with sharply rising climate damages, such as disintegration of ice sheets or large-scale release of permafrost carbon \citep{ArmstrongMcKay2022}. $h(\cdot)$ is a smooth, increasing, and convex function, and $h(x) \approx 0$ for $x \ll 0$, and $\zeta > 0$ is a coefficient scaling the magnitude of damages from tipping points. This smooth damage approximation is motivated by the tipping risks studied by \citet{LemoineTraeger2014} and \citet{CaiLontzek2019}, but does not introduce a separate irreversible tipping state.

$\Omega_{acid}(S)$ represents the cost of ocean acidification and other \textit{non-temperature} costs of atmospheric carbon. It cannot be masked by geoengineering and depends directly on the stock $S$. Estimates of damages from ocean acidification are unfortunately very limited. We choose $\Omega_{acid}(S)$ to add \$\CalAcidFloor/tCO$_2$ to the carbon price at current concentrations. The existence of $\Omega_{acid}(S)>0$ implies that the social cost of carbon remains positive even if geoengineering reduces standard climate damages to zero.

We assume quadratic temperature damages, $\omega(T)=\omega_2 T^2$, following DICE \citep{Nordhaus2017}. We use $h(x)=\max\{x,0\}^2$ and $\Omega_{acid}(S)=\omega_{acid}\max\{S-280,0\}^2$. We set $\omega_2=\CalOmegaTwo$, implying temperature-related output losses of approximately 1.8\% at 2$^\circ$C and 4.0\% at 3$^\circ$C.\footnote{Published SCC estimates vary substantially with discounting and damage assumptions. For example, \citet{Rennert2022} report roughly \$185/tCO$_2$ at a 2\% near-term rate and \$80/tCO$_2$ at 3\%. We also report a stronger-damage calibration informed by \citet{HowardSterner2017}.}

\begin{assumption}[Damage Regularity] \label{ass:damage_reg}
$\Omega$ is continuously differentiable, increasing, and convex, and twice continuously differentiable away from the tipping threshold $\bar{T}$. The calibrated tipping term is a quadratic hinge, so its second derivative steps at $\bar{T}$.
\end{assumption}

\subsection{Geoengineering and Ecological Debt} \label{subsec:geo_debt}

We assume the existence of a climate-vulnerable sovereign actor (Region V) capable of deploying geoengineering unilaterally. The credible threat of unilateral deployment may also induce other countries to coordinate deployment before the vulnerable country acts alone. In either case, deployment occurs at intensity $g_t \geq 0$, measured in degrees Celsius of cooling. Geoengineering artificially lowers the realized global temperature without altering the atmospheric carbon stock:
\begin{equation} \label{eq:temp_geo}
    T_t = \Psi(S_t) - g_t
\end{equation}

The vulnerable region $V$ in our model matches potential deployers in the real world: countries where climate damages may soon become so great that it becomes privately optimal to deploy SRM unilaterally. One motivating example involves wet-bulb temperature, a combined measure of heat and humidity. At wet-bulb temperatures around 35$^\circ$C, sustained exposure is considered an approximate limit beyond which even healthy people in shaded, ventilated environments suffer from deadly heat stress \citep{SherwoodHuber2010}. If a country faces repeated exposures to these conditions, it might choose to implement SRM rather than accept large numbers of heat-related fatalities. This possibility is empirically relevant because wet bulb temperatures have come close to this threshold in localized areas of South Asia and the Persian Gulf \citep{RaymondMatthewsHorton2020}, and projections identify the Indus and Ganges basins, the Persian Gulf, and other exposed regions as potential hotspots in the future \citep{PalEltahir2016,ImPalEltahir2017}. A climate-vulnerable country might also be pressured to deploy before reaching the wet-bulb survivability threshold. For example, governments could react early if they experience significant losses in agriculture or labor productivity due to heat and other climate changes.

In our model, $V$ should be interpreted as a reduced-form representation of countries for which the private value of geoengineering is unusually high, even if the global costs and risks of SRM are substantial.

\begin{assumption}[Geoengineering Policy] \label{ass:geo_policy}
The geoengineering intensity $g(S)$ is a known, exogenous policy function that is weakly increasing in the carbon stock: $g'(S) \geq 0$.
\end{assumption}

Our baseline calibration further assumes a specific rule for $g(S)$, hereafter referred to as the \textit{thermostat}:
\begin{equation} \label{eq:thermostat}
    g(S) = \max\left(0, \Psi(S) - T^*\right), \qquad T^* = \CalTstar^\circ\text{C}
\end{equation}
Our results state when they require the thermostat rather than the general form in Assumption~\ref{ass:geo_policy}. Under the thermostat, deployment activates when the unmitigated temperature reaches $T^*$, at $S = \Psi^{-1}(T^*) \approx \CalActivation$ ppm, and holds realized temperature constant at $T^*$ everywhere above the activation threshold. This rule is motivated by three considerations:

First, we set the activation threshold at $T^*=2^\circ$C as an illustrative point at which severe local climate damages could trigger deployment. Humid-heat studies show growing exposure in the Indus-Ganges basins and Persian Gulf, but do not identify a single global deployment threshold \citep{RaymondMatthewsHorton2020,ImPalEltahir2017,PalEltahir2016}. In our mapping, this temperature is reached at approximately \CalActivation\ ppm, \CalActivationGap\ ppm above the reference stock. We also examine earlier deployment.

Second, the intensity. Any geoengineering program that masks less than 100\% of incremental warming above $T^*$ is not credible, because realized temperature would drift above $T^*$ and Region V would unilaterally top the program up. The same argument disciplines whoever operates the program, including countries that coordinate deployment before the vulnerable country acts alone or successor coalitions that may take over geoengineering operations after V launches. The thermostat is the minimal deployment rule consistent with V's threat point. This is consistent with \citet{Weitzman2015}'s free-driver logic: the global thermostat is set by the most vulnerable actor with deployment capacity.

Third, exogeneity. We assume a global planner sets the carbon tax but takes the deployment rule $g(S)$ as given. This asymmetry is deliberate. Carbon taxation requires global coordination, while geoengineering is cheap enough that a vulnerable country can deploy it on its own. We maintain this asymmetry throughout the paper.\footnote{The globally coordinated carbon tax is itself an optimistic assumption that may not match current reality. In supplementary analysis, we estimate the model assuming the implemented carbon tax is a fraction of the social cost of carbon.}

In robustness analysis, we also study earlier and partial deployment. An announced program beginning at temperature $T_d$ offsets a share $m$ of subsequent warming until realized warming reaches $2^\circ$C, after which it prevents further warming:
\begin{equation} \label{eq:partial_rule}
 g(S)=\max\{0,m[\Psi(S)-T_d],\Psi(S)-2\}.
\end{equation}
For $m=1$, the program holds temperature at $T_d$. For $m=0.5$ and $T_d=1.5^\circ$C, it begins early and limits rather than immediately stops further warming, in the spirit of \citet{KeithMacMartin2015}.

We define the \textit{ecological debt} as:
\begin{equation} \label{eq:eco_debt}
    E(S) \equiv g(S)
\end{equation}
Ecological debt represents the gap between realized temperatures and the temperature the world would experience given the carbon stock in the absence of geoengineering. When $g = 0$, ecological debt is zero, so there is no masked warming. When $g > 0$, ecological debt tracks the carbon stock. Apart from carbon-cycle fluctuations, it rises when emissions exceed removal and falls when removal exceeds emissions. If deployment ever stops, the world experiences a termination shock as temperatures jump to the unmasked level. The shock is fast and severe because of a mismatch in timescales: carbon dioxide remains in the atmosphere for centuries, but stratospheric aerosols dissipate within one to two years, so temperature snaps back rather than rising gradually as in baseline climate change.

\begin{assumption}[Termination Hazard] \label{ass:term_hazard}
The termination hazard rate is:
\begin{equation} \label{eq:term_hazard}
    \lambda(S) = \bar{\lambda} \cdot \mathbf{1}\{g(S) > 0\}
\end{equation}
constant at $\bar{\lambda} > 0$ while a program is deployed and zero otherwise: with no program, there is nothing to terminate.
\end{assumption}

For simplicity, we assume the hazard rate is constant, so the probability of termination does not depend on the size of the program or the state of the climate. However, the damage from termination increases with the accumulated ecological debt. These assumptions capture the central termination mechanism: a positive probability of interruption and losses that increase with ecological debt. We will show that even though the physical hazard is flat, the market price of termination risk increases with the carbon stock. We calibrate $\bar{\lambda} = \CalLambdaBar\%$ per year, whose inverse is \CalLifetime\ years of continuous operation, and we report sensitivity analysis in lifetime units. We also consider a rising hazard $\lambda_0 + \lambda_1 g(S)$, reflecting the possibility that geoengineering at greater scale causes unforeseen harm to human health or ecosystems, forcing the program to stop. In the case of SAI, larger sulfur injections could create risks beyond those observed following volcanic eruptions. Finally, we consider a decreasing hazard with a positive floor, reflecting stronger efforts to prevent termination as its costs grow.

Termination in our model is absorbing: after a termination, geoengineering is never redeployed. We make this assumption mainly for tractability, but it may also be realistic. A program can end for many reasons, including political opposition, war, or technical accident. If opposition caused the halt, the same forces would stand in the way of redeployment. And whatever the cause, a world that has just suffered catastrophic losses may lack both the capacity and the political will to redeploy.

The permanent productivity loss upon termination is:
\begin{equation} \label{eq:Gamma}
    \Gamma(E) = 1 - \exp\!\left(-\Gamma_0 \, E^{\beta}\right)
\end{equation}
where $\Gamma_0 > 0$ scales the severity of termination damage and $\beta > 1$ determines its convexity. The functional form has three natural properties. If there is no ecological debt, termination causes no damage: $\Gamma(0) = 0$. Damage increases with the debt: $\Gamma' > 0$. And damage cannot exceed total productivity: $\Gamma(E) \to 1$ as $E \to \infty$. $\beta > 1$ implies that damage is convex at low levels of debt, so terminating a young program with little masked warming would do limited harm. However, terminating a mature program with a large accumulated debt would be catastrophic.

Aggregate productivity $A_t$ is constant except at termination:
\begin{equation} \label{eq:productivity}
    \frac{dA_t}{A_t} = -\Gamma(E_t) \, dN_t
\end{equation}
where $dN_t$ is a Poisson jump process with hazard rate $\lambda(S_t)$. When termination occurs ($dN_t = 1$), a fraction $\Gamma(E_t)$ of productivity is permanently destroyed. After termination, geoengineering ceases ($g = 0$, $\lambda = 0$) and the economy continues as a standard climate economy at the inherited carbon stock, with lower productivity. Climate damages, abatement costs, and the other costs in the model reduce consumption directly as proportional wedges on output (Section~\ref{subsec:planner}). We abstract from trend growth and other sources of productivity risk for simplicity.

Termination in our model is meant to represent a broad range of potential failure modes for geoengineering. A program may terminate because deployment causes (or is about to cause) negative side-effects, such as disruption of regional monsoons, damage to the ozone layer, or crop losses \citep{robock2009benefits, proctor2018estimating}, and the deployer or a global coalition chooses to stop. A geoengineering program may also end through misattribution and backlash. If a major weather disaster is blamed on geoengineering, political support can collapse whether or not the blame is accurate \citep{ParkerIrvine2018}. Other reasons for termination include wars, sanctions, or losing the ability to fund a program that needs constant operation. Regardless of the exact cause, termination means a catastrophic snapback in temperatures, as modeled in our paper.\footnote{A limitation of our model is that it does not capture the variation in potential losses associated with different types of termination. If termination is caused by real side effects, a program harmful enough to be shut down would also plausibly impose costs while operating, and our consumption measure omits them. Adding those costs would make geoengineering less attractive in every state. If termination follows from misattributed blame, no such costs exist, and the model is complete as it stands.}

\subsection{Firms and Dividends} \label{subsec:firms}

We consider Brown (fossil fuel) and Green (renewable energy) firms, a potential fusion supplier, and CDR (carbon dioxide removal). Instead of modeling detailed industry structure, we assume competitive energy supply with quadratic costs. Total energy demand is
\begin{equation} \label{eq:energy_market}
 Q(p)=\bar Q(p/\bar p)^{-\epsilon}=B+G+F,\qquad
 K_j(x)=c_jx+\tfrac12 k_jx^2,\quad j\in\{B,G,F\}.
\end{equation}
Here $p$ is the energy price in units of output before climate damages, $B$ and $G$ are Brown and Green production, and $F$ is fusion production, which is zero before a fusion breakthrough. We assume $0<\epsilon<1$, so energy demand falls less than proportionally as its price rises. For simplicity, we abstract from trend growth in energy demand, while allowing actual energy use to respond to prices.\footnote{The fixed benchmark $(\bar Q,\bar p)$ specifies energy demand at a reference price. Its scale is illustrative rather than fitted to current global emissions. We measure energy prices in units of output before climate damages and normalize Brown production to equal emissions in ppm per year.}

Let the damage function $d(S)=1-\Omega(T)-\Omega_{acid}(S)$ represent the fraction of output remaining after climate damages. The carbon tax $\tau$ is in consumption units. Competitive supply satisfies
\begin{equation} \label{eq:energy_supply}
 B=\left[\frac{p-c_B-\tau/d(S)}{k_B}\right]^+,\qquad
 G=\left[\frac{p-c_G}{k_G}\right]^+,\qquad
 F=\left[\frac{p-c_F}{k_F}\right]^+
\end{equation}
when fusion is available, where $[x]^+=\max(x,0)$. The market clears at $Q=B+G+F$. A higher carbon tax reduces fossil fuel profitability by raising production costs and shifting energy demand toward cleaner substitutes.

Firm dividends per unit of productivity are operating profits:
\begin{align}
 D^B &= d(S)[pB-K_B(B)]-\tau B=\tfrac12 d(S)k_B B^2, \label{eq:brown_dividend}\\
 D^G &= d(S)[pG-K_G(G)]=\tfrac12 d(S)k_G G^2. \label{eq:green_dividend}
\end{align}
Thus, zero Brown production means zero Brown dividends. Holding the technology set and $d(S)$ fixed, a higher carbon tax reduces Brown profits and raises Green profits. A fusion breakthrough changes the technology set: the new supplier can displace both Brown and Green energy. Its inventor receives a share $s_F$ of the fusion supplier's operating profits, with $s_F=1$ in the baseline, distinct from the profits of existing Green firms.

CDR firms operate carbon removal infrastructure, such as direct air capture plants, and receive government payments equal to the prevailing SCC for each ton removed. Section~\ref{subsec:innovation} describes how the value of the CDR patent depends on its profits after discovery.

After termination, all firms operate in an economy with permanently lower productivity. Previously discovered technologies remain available, and further research remains possible. Dividend flows are scaled by $(1-\Gamma(E))$ relative to the same post-termination economy without that productivity loss.

\subsection{The Planner's Problem} \label{subsec:planner}

The social planner maximizes the welfare of a representative global household with CRRA preferences, risk aversion $\gamma$, and discount rate $\rho > 0$:
\begin{equation} \label{eq:preferences}
    \mathbb{E}_0 \int_0^\infty e^{-\rho t} \, \frac{C_t^{1-\gamma}}{1-\gamma} \, dt
\end{equation}
Aggregate consumption equals output net of the costs the economy incurs:
\begin{equation} \label{eq:output}
 C_t=A_t c_t,\qquad c_t=d(S_t)\left[1-\Delta K_t+\Delta U_t\right]-c_R R_t.
\end{equation}
Here $\Delta K=\sum_j K_j(x_j)-\bar K$ is energy production cost relative to the fixed zero-tax, pre-fusion reference allocation, and $\Delta U=\int_{\bar Q}^{Q}p(q)\,dq$ is the benefit of additional energy relative to that allocation. This consumption-equivalent benefit allows energy demand to respond to prices without adding a state variable. Abatement occurs through substitution toward clean energy and lower total demand. Carbon removal uses resources $c_R R$, with $R=0$ before its breakthrough. Carbon taxes and payments for carbon removal are transfers within the economy, not additional resource costs. Prices, costs, and dividends are normalized per unit of productivity $A$. Dollar values use a pre-termination annual output benchmark of \$100 trillion. We retain private incentives for research but omit aggregate research spending from consumption as a simplifying approximation.

In our baseline calibration, we assume the global social planner sets the carbon tax $\tau_t$ equal to the Social Cost of Carbon (SCC), taking the geoengineering deployment rule $g(S)$ as given. The coordinated tax is an optimistic assumption: decentralized national carbon taxes would be lower than the global SCC, and carbon would accumulate faster. We explore this possibility in later extensions.

\subsubsection*{Homogeneity}

Because flow utility is CRRA and costs and rewards scale with productivity, the value function is homogeneous of degree $1-\gamma$ in productivity:
\begin{equation} \label{eq:separability}
    V(S, A) = A^{1-\gamma} \, v(S)
\end{equation}
where $v(S)$ solves the reduced HJB equation below. Under homogeneity, the planner's problem can be simplified to a one-dimensional ODE in $S$, and a permanent productivity loss of $\Gamma$ at termination enters the continuation value as the factor $(1-\Gamma)^{1-\gamma}$.

\subsubsection*{The HJB Equation: Incremental Construction}

We construct the HJB equation in three nested regimes, each adding one mechanism. We first omit breakthroughs and removal, then add them below. Here $u(c)=c^{1-\gamma}/(1-\gamma)$ represents flow utility per unit of $A^{1-\gamma}$. Given the carbon stock $S$ and tax $\tau$, competitive energy supply and market clearing determine Brown production $B(S,\tau)$ and normalized consumption $c(S,\tau)$ from Equation~(\ref{eq:output}).

\paragraph{Regime 1: No Geoengineering.} Set $g \equiv 0$, so $E = 0$, $\lambda = 0$, and $T = \Psi(S)$. This is the standard climate economy without geoengineering. The reduced HJB for $v(S)$ is:
\begin{equation} \label{eq:HJB_R1}
    \rho \, v = \max_{\tau} \left\{ u(c(S,\tau)) + B(S,\tau) \, v' + \tfrac{1}{2}\sigma_S^2 \, v'' \right\}
\end{equation}
where primes denote derivatives with respect to $S$. We set the carbon tax equal to the marginal welfare cost of emissions, expressed in consumption units. When Brown production is positive, this equality follows from the planner's first-order condition:
\begin{equation} \label{eq:SCC_R1}
    \tau = SCC^{R1}(S) = \frac{-v'(S)}{u'(c(S,\tau))}
\end{equation}
When Brown production is zero, further tax increases do not reduce emissions, so the maximization no longer uniquely determines the tax. We continue to set the carbon tax equal to the SCC, which also determines payments for carbon removal and the associated innovation incentives introduced below. In this regime, a higher carbon stock raises standard temperature damages and acidification damages, but does not affect termination damages.

\paragraph{Regime 2: Geoengineering Without Termination Risk.} Geoengineering is activated $g(S) > 0$, but termination risk is zero ($\lambda \equiv 0$). Realized temperature is $T = \Psi(S) - g(S)$. The HJB becomes:
\begin{equation} \label{eq:HJB_R2}
    \rho \, v = \max_{\tau} \left\{ u(c(S,\tau)) + B(S,\tau) \, v' + \tfrac{1}{2}\sigma_S^2 \, v'' \right\}, \qquad T = \Psi(S) - g(S)
\end{equation}
The difference between Regimes 1 and 2 is that temperature damages $\Omega$ are evaluated at the artificially low realized temperature rather than at $\Psi(S)$. Acidification damages do not differ from Regime 1 because they depend directly on the carbon stock. This decline in temperature damages reduces the planner's incentive to tax carbon (the moral hazard channel).

\paragraph{Regime 3: The Full Model.} Activate $\lambda(S) > 0$ and $\Gamma(E) \, dN_t$. The HJB acquires a jump term reflecting the expected value loss from termination:
\begin{equation} \label{eq:HJB_R3}
\begin{split}
    \rho \, v = \max_{\tau} \Big\{ & u(c(S,\tau)) + B(S,\tau) \, v' + \tfrac{1}{2}\sigma_S^2 \, v'' \\
    & + \lambda(S) \left[ (1-\Gamma(E(S)))^{1-\gamma} \, v^{R1}(S) - v \right] \Big\}
\end{split}
\end{equation}
The new term is the expected value change from a termination event. After termination, the economy moves immediately to Regime 1 with the same carbon stock and permanently lower productivity. The continuation value is the Regime 1 value $v^{R1}(S)$, scaled by the factor $(1-\Gamma(E))^{1-\gamma}$. Under this setup, the world after deployment faces a small chance each instant of a large permanent loss, and this possibility lowers value throughout deployment.

\subsubsection*{The SCC in Regime 3}

In Regime 3, the social cost of carbon reflects both ongoing climate damages and potential losses from termination. Under the thermostat policy, realized temperature is held constant at $T^*$ during deployment, so an additional unit of carbon causes no additional current temperature damage, although it still increases acidification damages. It also raises ecological debt $E(S)$ and the damage $\Gamma(E)$ if termination occurs:
\begin{equation} \label{eq:SCC_full}
 SCC^{R3}(S)=-v'(S)/u'(c(S,\tau)).
\end{equation}
The slope of the value function captures these effects and their implications for future welfare. When we introduce technological breakthroughs below, it also incorporates their effects on future damages and policy.

We solve the HJB numerically on a grid of carbon stocks, iterating between policy choices and their implied welfare values until convergence (Section~\ref{sec:numerical}). The SCC is forward looking: it accounts for future carbon accumulation, deployment, termination risk, and technological breakthroughs. We continue to set the carbon tax and payment for removal equal to the SCC even when it exceeds the tax needed to shut down Brown production. At zero emissions, the marginal harm from carbon determines the payment for carbon removal.

\subsection{Breakthroughs and Removal Capacity} \label{subsec:breakthroughs}

Two technologies can arrive as Poisson events, with arrival rates that depend on the carbon stock, regime and available technologies:

A fusion breakthrough $N_{F,t}$ introduces a low-cost clean energy supplier with cost function $K_F$. Fusion competes with both Brown and existing Green energy. Fusion is meant to represent a new low-cost form of clean energy. It does not describe incremental improvements in solar and wind already represented by Green energy supply. We calibrate cost parameters for fusion to target a 90\% emissions reduction at a \$75/tCO$_2$ tax in the fixed-demand reference economy.

A carbon removal breakthrough $N_{C,t}$ allows for cheap removal at scale. We assume a CDR breakthrough makes capacity $\bar R=\CalRbar$ GtCO$_2$/yr (\CalRbarPpm\ ppm/yr) available at negligible operating cost, $c_R=0$. This stylized assumption captures an important limitation: even a breakthrough that substantially lowers carbon removal costs would not allow us to draw down accumulated carbon immediately, because annual capacity remains limited. Even at maximum capacity, removing \CalRbarPpm\ ppm per year takes decades to unwind a century of buildup. Our assumed capacity limit is of the order considered in long-run portfolio assessments \citep{NASEM2019}, while our assumption of negligible cost is deliberately favorable to removal. For simplicity, we also assume that full capacity is available immediately upon discovery, although a real removal industry may require additional time to build capacity.

After the CDR breakthrough, the planner chooses removal $R\in[0,\bar R]$, and the government pays operators the SCC for each ton removed. Our baseline assumes zero operating costs so that we can focus on limited capacity as the main friction. Removal therefore runs at full capacity whenever the SCC is positive. Even at full capacity, removal takes time to reduce accumulated carbon, and new emissions can offset it. Section~\ref{subsec:fig_procurement} shows this balance.

\paragraph{Safe landing.} We define a safe landing as reaching, after geoengineering deployment and before termination, carbon levels such that $g(S)=0$ and equilibrium net emissions $B-R$ are nonpositive. Under the thermostat rule, this means returning to $S\leq\CalActivation$ ppm and switching off geoengineering. We do not require any particular technological breakthrough to reach safe landing by definition, but we will show that both breakthroughs are needed in our baseline calibration.

\subsection{Innovation} \label{subsec:innovation}

We model innovation so that it responds to financial incentives. We adopt a reduced form instead of modeling researchers' decisions directly. Each breakthrough has a background arrival rate and a component that increases with the market value of the patent its innovator would receive, up to a cap:
\begin{align} \label{eq:innovation_hazard}
 \eta_F(S,z)&=\min\{a_F+\kappa_F\max(\hat\pi_F(S,z),0),\eta_{max}\},\\
 \eta_C(S,z)&=\min\{a_C+\delta_F\mathbf1\{F\in z\}+\kappa_C\max(\hat\pi_C(S,z),0),\eta_{max}\}. \nonumber
\end{align}
Here $z$ records the technologies already available; each rate applies only before its own discovery. We suppress $z$ and the regime index where the state is clear. The positive background rates represent progress that is less sensitive to market rewards, such as low-cost artificial intelligence innovation abilities. The additional CDR rate $\delta_F$ after fusion captures the opening of a route to removal technologies that relies on cheap energy. The reward-sensitive component can be motivated by a micro-founded model with quadratic private research costs. We retain these incentives but omit aggregate research spending from consumption for tractability.

The patent value $\hat\pi_i(S,z)$ is what the innovator's claim would be worth immediately after discovery, per unit of productivity $A$. Like other assets, it is priced in the post-discovery equilibrium using the jump-adjusted valuation equation.

The fusion inventor receives a share $s_F=1$ of the new supplier's operating profits in the baseline, $d(S)k_F F^2/2$. The CDR inventor receives operating profits $(\tau-c_R)R$ after discovery. Governments offer no additional innovation prize. Carbon removal can lower the carbon price that rewards its inventor, even when operating costs are negligible.

Geoengineering can weaken innovation incentives through lower future energy and carbon prices. Termination risk partly reverses this effect by making emissions reduction and removal more valuable, but termination also destroys productivity. Importantly, the impact of geoengineering and termination risk on fusion and CDR incentives can therefore differ. Fusion profits depend on energy demand and competition from existing suppliers. CDR profits depend on the carbon price, which reflects the value of reducing ecological debt. Fusion can reduce new emissions but cannot remove the carbon already accumulated. We solve for both patent values, allowing either to rise or fall at termination.

Finally,  we impose a cap of $\eta_{max}=\CalEtaMax\%$ per year. This bounds the expected wait below by \CalEtaMaxYears\ years, without imposing a minimum realization time. Both technologies remain available after discovery, and research toward any remaining breakthrough continues after termination.

\subsection{Solving the Full Model} \label{subsec:prearrival_hjb}

Regime 3 consists of eight economies, one for each combination of technology state (neither breakthrough, fusion only, removal only, or both) and termination status. Regime 2 adds four technology states without termination risk. We also solve the four Regime 1 states, both as the benchmark without geoengineering and to obtain the continuation values after termination in Regime 3, adjusted for the permanent productivity loss. Each random event (termination or breakthrough) moves the world from one economy to another. Each economy is an optimizing problem of the same form as Section~\ref{subsec:planner}, with the economies it can jump to as continuation values. We solve by backward induction, starting from the economies that no further event can leave. No circularity arises, because nothing that happens after a breakthrough depends on the taxes, values, or research intensities that prevailed before it.

The post-termination economies follow Regime 1 policies for the same technology state, with permanently lower productivity and no possibility of redeploying geoengineering. Available technologies remain in use, and research toward any remaining breakthrough continues. Within each regime, we solve the four technology states in the following order:
\begin{enumerate}
    \item \textit{The economy with both breakthroughs}, $v_B$. Removal is paid the SCC. In Regime 3, the termination hazard remains while $g>0$.
    \item \textit{The economy with removal only}, $v_C$. Removal can lower the carbon stock, but residual emissions may offset it. A fusion breakthrough moves the economy to $v_B$.
    \item \textit{The economy with fusion only}, $v_F$. Cheap energy reduces new emissions but does not remove accumulated carbon. A CDR breakthrough moves the economy to $v_B$.
    \item \textit{The economy before either breakthrough}, solved last within each regime, with arrival rates derived from the corresponding post-discovery patent values.
\end{enumerate}

The last step of the backward induction is the pre-breakthrough problem in item 4. The Regime 3 problem adds two jump terms to Equation~(\ref{eq:HJB_R3}), one for each breakthrough:
\begin{equation} \label{eq:HJB_full}
\begin{split}
    \rho \, v = \max_{\tau} \Big\{ & u(c(S,\tau)) + B(S,\tau) \, v' + \tfrac{1}{2}\sigma_S^2 \, v'' + \lambda(S) \left[ (1-\Gamma(E(S)))^{1-\gamma} \, v^{post}(S) - v \right] \\
    & + \eta_F(S) \left[ v_F(S) - v \right] + \eta_C(S) \left[ v_C(S) - v \right] \Big\}
\end{split}
\end{equation}
If a breakthrough occurs, the economy jumps to the corresponding continuation value defined above, $v_F$ or $v_C$. When choosing the current tax, the planner takes the breakthrough rates $\eta_i(S)$ as given. These rates depend on patent values in the post-discovery economies, whose policies are solved first. After termination, the economy follows Regime 1 policies with the same available technologies and research continuing. Thus, $v^{post}(S)$ equals the corresponding Regime 1 value, as in Equation~(\ref{eq:HJB_R3}). Each regime's breakthrough rates depend on patent values after discovery in that regime.

Setting $a_F=a_C=\delta_F=\kappa_F=\kappa_C=0$ shuts down innovation entirely, leaving a world in which no breakthrough will ever occur and today's technologies are permanent. We call this special case the \textit{no-innovation benchmark}. Comparing the full model against the no-innovation benchmark allows us to see what endogenous innovation adds.

\subsection{Asset Pricing} \label{subsec:asset_pricing}

\subsubsection*{The Stochastic Discount Factor}

Under CRRA utility, the stochastic discount factor evolves as:
\begin{equation} \label{eq:SDF}
    \frac{d\Lambda_t}{\Lambda_t} = -r_t \, dt - \gamma \xi_t \, dZ_t + \sum_{j \in \{term, F, C\}} \left[ (J^{\Lambda}_{j,t} - 1) \, dN_{j,t} - h_{j,t} (J^{\Lambda}_{j,t} - 1) \, dt \right]
\end{equation}
where the sum has three terms, one for each event that can cause a jump in consumption: termination (hazard $h_{term} = \lambda$), a fusion breakthrough ($h_F = \eta_F$), and a CDR breakthrough ($h_C = \eta_C$). The coefficient $\xi_t$ measures how strongly consumption responds to the random carbon-stock fluctuations in Equation~(\ref{eq:carbon_stock}), while $\gamma$ determines how strongly investors value the resulting changes in consumption. Since these carbon stock fluctuations are small, we set $\xi_t = 0$ below. In the no-innovation benchmark, only the termination term is present. For each event, the jump in the pricing kernel is the marginal utility ratio across the event:
\begin{equation} \label{eq:SDF_jump}
    J^{\Lambda}_{j,t} = \left(\frac{C_{t+}}{C_{t-}}\right)^{-\gamma}
\end{equation}
At termination, productivity is destroyed by the factor $1-\Gamma(E_t)$, and renewed warming and changed energy allocation affect normalized consumption. Let $J^c_t=c^{post}(S_t,z_t)/c(S_t,z_t)$ represent the latter change. Then $C_{t+}/C_{t-}=(1-\Gamma(E_t))J^c_t$. The sign of the pricing-kernel jump depends on this total consumption change.

\subsubsection*{Risk-Neutral Pricing}

The change of measure to the risk-neutral probability $\mathbb{Q}$ scales each physical hazard by its pricing-kernel jump. For termination:
\begin{equation} \label{eq:lambda_Q}
    \lambda^{\mathbb{Q}}_t = \lambda_t \cdot J^{\Lambda}_t = \lambda_t \cdot \left[(1 - \Gamma(E_t)) \cdot J^{c}_t\right]^{-\gamma}
\end{equation}
    
This approach applies the rare-disaster mechanism in \citet{Barro2006} to the termination shock. When termination lowers consumption, marginal utility rises, so termination is priced as if it were more likely than it actually is. In our baseline calibration, the physical hazard is constant while a program is deployed. A larger consumption loss at termination therefore raises the risk-neutral hazard even without any increase in the physical hazard.

The same change of measure applies to the two potential breakthroughs. The risk adjustment depends on their immediate effect on consumption, which need not have the same sign as their effect on lifetime welfare. Fusion adds a low-cost energy supplier. CDR changes carbon prices and energy allocation. The risk-neutral hazard is adjusted downward when consumption rises and upward when it falls. We discuss the signs with the results in Section~\ref{subsec:fig_trp_decomp}.

\subsubsection*{Valuation Equations}

As noted above, we approximate the compensation for carbon-stock fluctuations as zero, $\xi_t=0$, and use a constant asset discount rate $r$. We retain the carbon-stock diffusion ($\sigma_S = \CalSigmaS$ ppm/$\sqrt{\text{yr}}$) but approximate its price of risk as zero, leaving the risk adjustment to jumps.\footnote{The constant discount rate and zero price of diffusion risk are pricing approximations. We do not solve for the endogenous risk-free rate implied by CRRA utility. Allowing diffusion risk to be priced would add the covariance term $-\gamma \xi_t \sigma_S V^{i\prime}(S)$ to Equation~(\ref{eq:VB_R3}).} Each firm's value then solves an ODE on the carbon-stock grid that accounts for dividends, changes in the carbon stock, and possible termination or breakthroughs. These events enter at their risk-neutral arrival rates, $h_j^{\mathbb{Q}}(S) = h_j(S)J_j^\Lambda(S)$, where $h_{term}=\lambda$, $h_F=\eta_F$, and $h_C=\eta_C$ are the physical arrival rates of termination, a fusion breakthrough, and a CDR breakthrough, respectively.

\begin{equation} \label{eq:VB_R3}
    r V^i(S) = D^i(S,z) + \mu_S(S) V^{i\prime}(S) + \tfrac{1}{2}\sigma_S^2 V^{i\prime\prime}(S) + \sum_j h^{\mathbb{Q}}_j(S) \left[ V^i_j(S) - V^i(S) \right]
\end{equation}
Here, $\mu_S(S)$ is the equilibrium drift of the carbon stock and $V^i_j(S)$ is firm $i$'s value in the economy that event $j$ leads to. At termination, this value is scaled by the productivity loss, $V^i_{post}(S) = (1-\Gamma(E(S))) \cdot V^{i,R1}(S)$.

The fusion and CDR patent values solve the same equation using their operating profits after discovery. Their carbon dynamics, future technologies and termination destinations are those of the corresponding post-discovery economy. This ensures that innovation rewards and welfare follow the same continuation rules.

In Regimes 1 and 2, the termination jump is zero ($\lambda = 0$). In the no-innovation benchmark the breakthrough jumps are zero as well, and asset values are the discounted present values of their future dividend flows.

\subsubsection*{Jump Risk Premia}

The expected excess return of any firm $i$ satisfies:
\begin{equation} \label{eq:expected_return}
    \mu^i - r =
    \underbrace{\gamma \xi \, \sigma_{i,Z}}_{\text{Diffusion premium } (= 0 \text{ here})}
    \;\underbrace{-\sum_{j \in \{term,F,C\}} h_j
    \left(J^{\Lambda}_j - 1\right)\left(J^i_j - 1\right)}_{\text{Jump risk premium}}
\end{equation}
where $\sigma_{i,Z}=\sigma_S V^{i\prime}/V^i$ is the return exposure to carbon-stock fluctuations and $J^i_j \equiv V^i_j/V^i$ is the jump multiplier for firm $i$'s valuation at event $j$. The termination component represents compensation investors demand for bearing termination risk. Termination lowers consumption. Firms that lose value at termination ($J^i < 1$) covary negatively with the jump in the stochastic discount factor, so they have a positive risk premium. Firms that gain value at termination ($J^i > 1$) have a negative risk premium. Likewise, the two breakthrough components price exposure to breakthrough news (the full decomposition details are reported in Section~\ref{subsec:fig_trp_decomp}).

\subsubsection*{Three Nested Regimes}

In summary, we analyze the economy under three nested parameter restrictions, each isolating a distinct economic mechanism:

\begin{center}
\small
\setlength{\tabcolsep}{5pt}
\begin{tabular}{lccc}
\toprule
& Regime 1 (No geo.) & Regime 2 (Geo, no term.) & Regime 3 (Full) \\
\midrule
$g(S)$ & $\equiv 0$ & $\geq 0$ (Assumption~\ref{ass:geo_policy}) & $\geq 0$ (Assumption~\ref{ass:geo_policy}) \\
$\lambda(S)$ & $\equiv 0$ & $\equiv 0$ & $\geq 0$ (Assumption~\ref{ass:term_hazard}) \\
$\Gamma(E) \, dN_t$ & absent & absent & active \\
Arrivals $N_F, N_C$ & live & live & live \\
\bottomrule
\end{tabular}
\end{center}

In all three regimes, both breakthroughs can occur. The rates at which they occur, $\eta_F(S)$ and $\eta_C(S)$, depend on each regime's patent values (Section~\ref{subsec:innovation}), so a change in carbon and energy prices can alter research incentives. Setting $a_F=a_C=\delta_F=\kappa_F=\kappa_C=0$ turns discovery off, and represents our no-innovation benchmark. This completes the model. We now turn to the theoretical results.

\section{Theoretical Results} \label{sec:results}

\subsection{Carbon Prices and Firm Valuations} \label{subsec:regime1} \label{subsec:regime2} \label{subsec:regime3}

A termination event moves the Regime 3 economy to Regime 1 with the same carbon stock and technology state, with permanently lower productivity. Firm $i$'s value after termination is therefore $(1-\Gamma(E))V^{i,R1}$. Comparing this with its value before termination determines whether it gains or loses at the event and, when consumption falls, the sign of its termination risk premium.

\begin{samepage}
\begin{corollary}[Firm Value and Termination Exposure] \label{prop:toxic_put}
At a deployed state with positive firm values, a positive termination hazard, and a fall in consumption at termination, firm $i$ has a positive termination risk premium if and only if
\begin{equation} \label{eq:termination_value_threshold}
 V^{i,R3}>(1-\Gamma(E))V^{i,R1}.
\end{equation}
In particular, a firm that is worth more with geoengineering than without it must lose value at termination and carry a positive termination premium. A firm that is already worth less than in Regime 1 can either gain or lose at termination.
\end{corollary}
\end{samepage}

\textit{Proof.} Substitute $J^i_{term}=(1-\Gamma(E))V^{i,R1}/V^{i,R3}$ into the termination component of Equation~(\ref{eq:expected_return}). Since $J^\Lambda_{term}>1$, the premium is positive exactly when $J^i_{term}<1$. \qed

In our calibration, geoengineering initially raises Brown value by suppressing carbon taxation. As ecological debt accumulates, the carbon price rises and Brown value eventually falls below its Regime 1 level. However, being worth less than in a world without geoengineering does not necessarily mean that the firm loses value at termination. At sufficiently high stocks, termination can lower the carbon tax enough to restart Brown production, raising equity value.

\subsection{Innovation Incentives} \label{subsec:innovation_results}

Geoengineering lowers realized temperatures, buying time for a fusion or carbon removal breakthrough. This is the peak-shaving argument of \citet{KeithMacMartin2015}. But deployment can also weaken innovation incentives, because innovators earn profits tied to energy and carbon prices. The two technologies earn these rewards in very different ways.

\begin{proposition}[Energy Profits and Removal Rewards] \label{prop:crowding_out}
After fusion is available, let $p_c$ be the energy price at which Green and fusion supply meet demand without Brown production, and define
\[
 M_F=\frac{s_F}{2k_F}\bigl([p_c-c_F]^+\bigr)^2.
\]
The fusion inventor's profits per unit of productivity are at most $d(S)M_F$. They reach this ceiling when Brown production is zero, and further increases in the carbon price do not raise them. With $0<d(S)\leq1$, non-growing productivity, and the constant discount rate $r$ in Equation~(\ref{eq:VB_R3}), its patent value satisfies $\hat\pi_F\leq M_F/r$.

By contrast, a discovered CDR technology with zero operating costs earns $\tau\bar R$ whenever the SCC is positive. Its removal reward continues to increase with the SCC beyond the carbon price that shuts down Brown production.
\end{proposition}

\textit{Proof.} Nonnegative Brown energy supply implies that the market-clearing energy price cannot exceed $p_c$. Substituting this price into fusion's supply and profit functions gives the operating-profit ceiling. All future profit flow is limited by the same ceiling, and termination cannot increase productivity. Discounting that ceiling at the constant rate $r$ gives the patent-value bound. When Brown production is zero, the energy market clears using only clean supply (Green or fusion) and no longer responds to the carbon tax. The removal reward follows from payment at the SCC and full-capacity operation at zero cost. \qed

This proposition explains why rising carbon prices can continue to reward carbon removal even after they cease to raise clean-energy profits. 

\subsection{Buying Time and Safe Landing}

A safe landing requires returning to the deployment threshold level of carbon, with nonpositive net emissions before termination occurs. Technological breakthroughs alone are insufficient to end a deployed geoengineering program. Fusion can halt further accumulation (aside from natural carbon fluctuations) but cannot remove the carbon already in the atmosphere.  A CDR breakthrough also cannot ensure a safe landing, because capacity limits imply that removal takes time.

\begin{samepage}
\begin{proposition}[Limited Capacity and Safe Landing] \label{prop:siege}
Under the thermostat rule, let $S_a=\Psi^{-1}(T^*)$ be the deployment threshold and suppose that the physical termination hazard is at least $\lambda_{\min}>0$ while $S>S_a$. With Brown emissions $B\geq0$ and removal bounded by $\bar R>0$, the safe landing probability from any deployed state satisfies
\begin{equation} \label{eq:landing_capacity_bound}
 p_z(S)\leq\exp\!\left[-\frac{2\lambda_{\min}(S-S_a)}{\sqrt{\bar R^2+2\lambda_{\min}\sigma_S^2}+\bar R}\right],\qquad S>S_a.
\end{equation}
This bound holds in every technology state $z$ and allows for subsequent discoveries and endogenous policy responses. Without carbon-stock noise, it reduces to $\exp[-\lambda_{\min}(S-S_a)/\bar R]$.
\end{proposition}
\end{samepage}

\textit{Proof.} With zero emissions and removal at full capacity, the carbon stock would follow $S-\bar R t+\sigma_S Z_t$. For any given sequence of random carbon-stock fluctuations, the actual stock cannot fall below this path, because emissions are nonnegative and removal cannot exceed capacity. Reaching the deployment threshold therefore cannot occur earlier than under this fastest possible drawdown. A safe landing requires surviving at least until that first crossing. The accumulated termination hazard before the crossing is at least $\lambda_{\min}$ times the elapsed time. Hence the landing probability is at most $\mathbb E[\exp(-\lambda_{\min}H)]$, where $H$ is the fastest process's first crossing time. This expectation is represented by the right side of Equation~(\ref{eq:landing_capacity_bound}). \qed

Even with immediate discovery of both breakthroughs and zero emissions, limited removal capacity leaves the world exposed to termination risk while carbon is being drawn down. Without carbon-stock fluctuations, returning from $S$ to the deployment threshold $S_a$ takes at least $(S-S_a)/\bar R$ years.

\section{Quantitative Results} \label{sec:numerical}

We solve the planner's HJB equations by policy iteration and the valuation equations by finite differences. The baseline uses 4,485 grid points over $S\in[280,\CalGridMax]$ ppm. All three regimes use the same parameter values, varying only the geoengineering and termination channels.

\subsection{Calibration} \label{subsec:calibration}

Table~\ref{tab:calibration} presents the parameter values. The climate, damage, and acidification parameters were discussed in Section~\ref{sec:model}. Risk aversion $\gamma=\CalRiskAversion$ is within the range of the rare disaster literature \citep{Barro2006,Wachter2013}. We calibrate the pre-fusion supply curves so that a carbon tax of \$\CalReferenceShutdown/tCO$_2$ eliminates Brown production when total energy demand is held fixed, allowing costly substitution away from Brown energy \citep{IPCC2022,GillinghamStock2018}. With fusion and the same fixed demand, a carbon tax of \$75/tCO$_2$ reduces Brown emissions by 90\%. Note that these are only calibration targets; in the model, energy demand responds to prices, so the tax needed to achieve either reduction can differ. The physical termination hazard is \CalLambdaBar\% per year while deployed.

We set the scale of the termination damage function $\Gamma(E)$ by specifying the loss at one reference point: termination at 800 ppm, with ecological debt of \SevDebtAtEightZeroZero$^\circ$C, destroys \CalGammaTargetEightHundred\% of productivity. We view this as a pessimistic assumption because it imposes a substantial permanent productivity loss in addition to the damages from renewed warming, although the magnitude of termination losses is highly uncertain. Evidence on warming and economic growth \citep{BurkeHsiangMiguel2015}, and on the ecological consequences of rapid termination warming \citep{TrisosCriscitielloNicholls2018}, motivates allowing additional losses but does not offer magnitude estimates. With $\beta=\CalBeta$, the target implies $\Gamma_0=\CalGammaZero$. The loss is zero at zero cooling and rises smoothly, reaching \SevLossAtSixZeroZero\% at 600 ppm and \SevLossAtSevenZeroZero\% at 700 ppm.

\begin{table}[htbp]
\centering
\caption{Baseline Calibration}
\label{tab:calibration}
\footnotesize
\setlength{\tabcolsep}{3pt}
\begin{tabular}{lll}\toprule Parameter & Symbol & Value \\ \midrule
Current carbon stock & $S_0$ & 427 ppm \\
Temperature response per doubling & $T_{cs}$ & 2.3$^\circ$C \\
Deployment temperature & $T^*$ & 2$^\circ$C \\
Masking share & $m$ & 1 \\
Tipping-damage threshold & $\bar T$ & 1.5$^\circ$C \\
Quadratic temperature damage & $\omega_2$ & 0.004394 \\
Tipping-damage coefficient & $\zeta$ & 0.005 \\
Quadratic acidification coefficient & $\omega_{acid}$ & 6.39534e-08 \\
Carbon-stock volatility & $\sigma_S$ & 1 ppm/$\sqrt{\mathrm{yr}}$ \\
Termination hazard & $\bar\lambda$ & 0.5\%/yr \\
Loss at 800 ppm & $\Gamma(E(800))$ & 12.5\% \\
Termination damage scale & $\Gamma_0$ & 0.0498 \\
Termination convexity & $\beta$ & 2.5 \\
Risk aversion & $\gamma$ & 8 \\
Welfare discount rate & $\rho$ & 2\% \\
Asset discount rate & $r$ & 4\% \\
Reference total energy & $\bar Q$ & 3.75 \\
Reference zero-tax brown emissions & $\bar e$ & 3 ppm/yr \\
Supply curvature parameter & $\psi=1/(k_B+k_G)$ & 120 \\
Reference green share & $\bar G/\bar Q$ & 20\% \\
Brown curvature share & $k_B/(k_B+k_G)$ & 0.15 \\
Fusion curvature ratio & $k_F/k_G$ & 1 \\
Reference energy price & $\bar p$ & 0.028698 \\
Demand elasticity & $\epsilon$ & 0.3 \\
Energy cost intercepts & $(c_B,c_G,c_F)$ & (0.024948, 0.023385, 0.014538) \\
Energy cost slopes & $(k_B,k_G,k_F)$ & (0.001250, 0.007083, 0.007083) \\
Reference energy cost & $\bar K$ & 0.1 \\
Fusion inventor profit share & $s_F$ & 1 \\
Background arrival rates & $a_F,a_C$ & 0.2\%/yr each \\
CDR uplift after fusion & $\delta_F$ & 1\%/yr \\
Fusion reward slope & $\kappa_F$ & 0.05207 \\
CDR reward slope & $\kappa_C$ & 0.05677 \\
Mean discovery waits from today & Fusion, CDR & 40, 50 years \\
Breakthrough-rate cap & $\eta_{max}$ & 25\%/yr \\
Removal cost & $c_R$ & \$0/tCO$_2$ \\
Removal capacity & $\bar R$ & 15 GtCO$_2$/yr \\
\bottomrule\end{tabular}

\end{table}

For the technological breakthroughs, we set common background arrival rates to 0.2\% per year and a one-percentage-point increase in the CDR rate after the arrival of a fusion breakthrough. The positive research slopes, $\kappa_F=\CalKappaF$ and $\kappa_C=\CalKappaC$, are set to jointly match mean waiting times of 40 years for fusion and 50 years for CDR from today's stock, $S_0=\CalSzero$ ppm, in Regime 3. These means include the other breakthrough and continued innovation after termination. 

We choose these mean waiting times to approximately match estimates of when these breakthroughs could occur in the literature. Fusion assessments distinguish initial commercial plants from a substantial energy-market contribution, with the latter generally placed in the 2050s and beyond \citep{EuropeanCommission2023Fusion,UKGOScience2025Fusion}. Our fusion event represents cheap competing supply at scale. The CDR event is more ambitious than current projections: an expert elicitation places median direct-air-capture capacity in 2050 at 1.69 GtCO$_2$ under strong climate policy, far below our 15 GtCO$_2$ capacity, and at substantially positive costs \citep{ShayeghEtAl2021}. We vary the timing and background arrival assumptions in the heterogeneity analysis.

\begin{figure}[htbp]
    \centering
    \includegraphics[width=\textwidth]{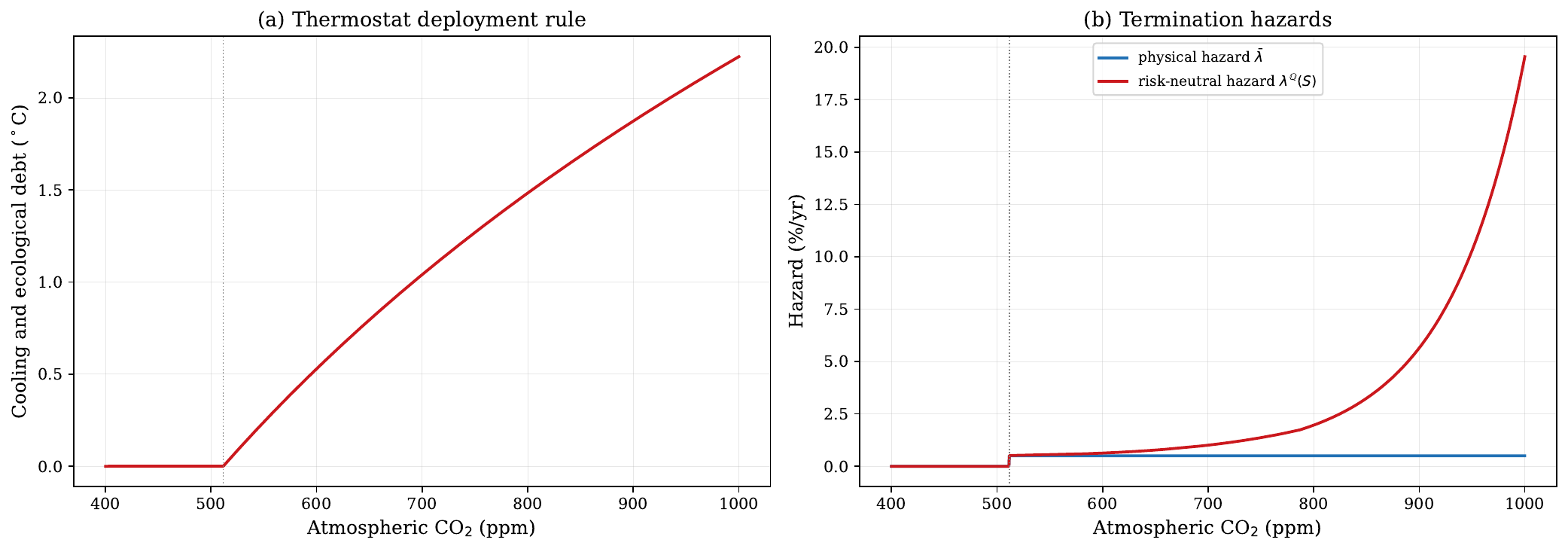}
    \caption{\textbf{Calibrated Model Primitives.} (a) Cooling and ecological debt under the thermostat rule. (b) Physical and risk-neutral termination hazards, conditional on neither breakthrough having arrived.}
    \label{fig:primitives}
\end{figure}

Figure~\ref{fig:primitives} shows the calibrated geoengineering policy. Panel (a) combines cooling and ecological debt, which coincide under the thermostat rule. Deployment begins at \CalActivation\ ppm. Panel (b) shows the constant physical termination hazard and the increasing risk-neutral hazard. The latter reflects the growing consumption loss conditional on termination.

\subsection{The Carbon Price and Emissions} \label{subsec:fig_scc} \label{subsec:fig_dynamics}

\begin{figure}[htbp]
    \centering
    \includegraphics[width=\textwidth]{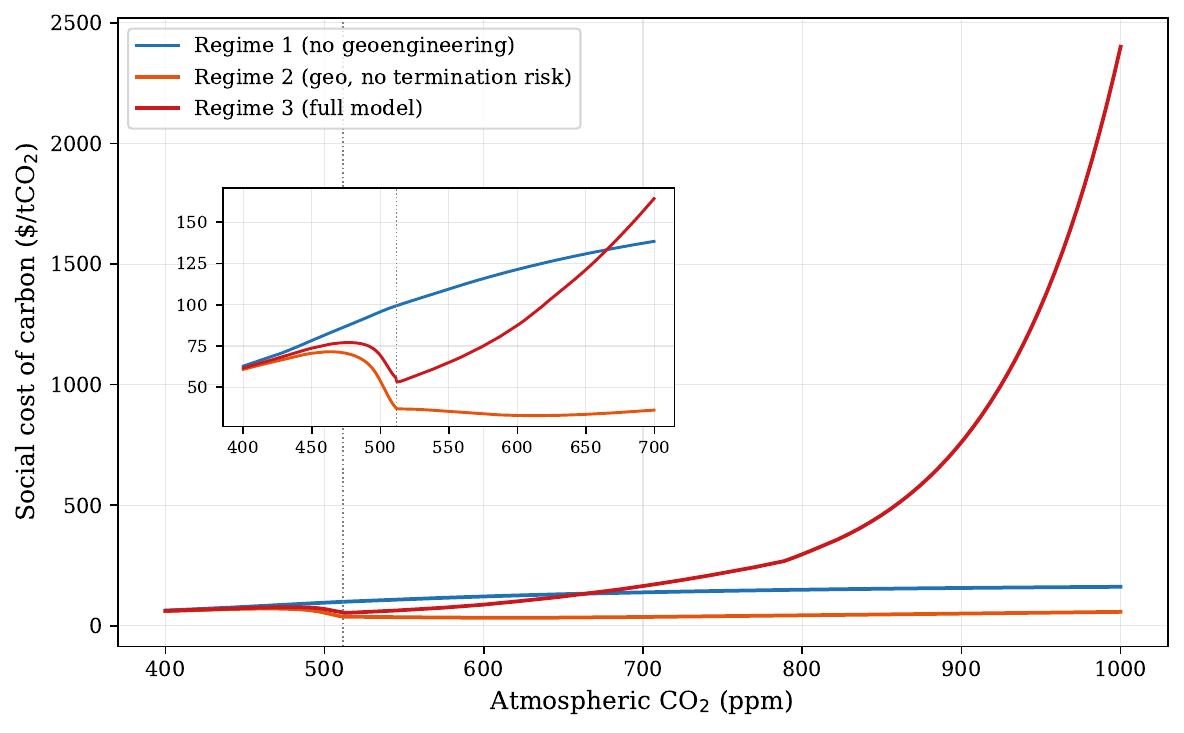}
    \caption{\textbf{Social Cost of Carbon.} Carbon prices before either breakthrough and before termination, accounting for future events. The Regime 3 price crosses above Regime 1 at $\hat S\approx\InnShat$ ppm.}
    \label{fig:scc}
\end{figure}

Figure~\ref{fig:scc} plots the social cost of carbon under each of the three regimes, conditional on neither breakthrough having yet arrived and, in Regime 3, before termination. The curves account for future events. In Regime 1 the price rises from \$\SCCLiveROneAtFourTwoSeven/tCO$_2$ at current concentrations to \$\SCCLiveROneAtSixZeroZero/tCO$_2$ at 600 ppm and \$\SCCLiveROneAtEightZeroZero/tCO$_2$ at 800 ppm.

In Regime 2, geoengineering suppresses marginal temperature damages and lowers the carbon price to \$\SCCLiveRTwoAtSixZeroZero/tCO$_2$ at 600 ppm. Termination risk partly restores the price in Regime 3. The price declines as deployment approaches, then rises as ecological debt accumulates. It crosses above the Regime 1 price at $\hat S\approx\InnShat$ ppm. Above that point, additional carbon is more socially costly with geoengineering than without it.

Appendix~\ref{app:dynamics} shows illustrative carbon-stock paths over time, with breakthrough and termination events suppressed.

\subsection{The Toxic Put} \label{subsec:fig_brown}

\begin{figure}[htbp]
    \centering
    \includegraphics[width=\textwidth]{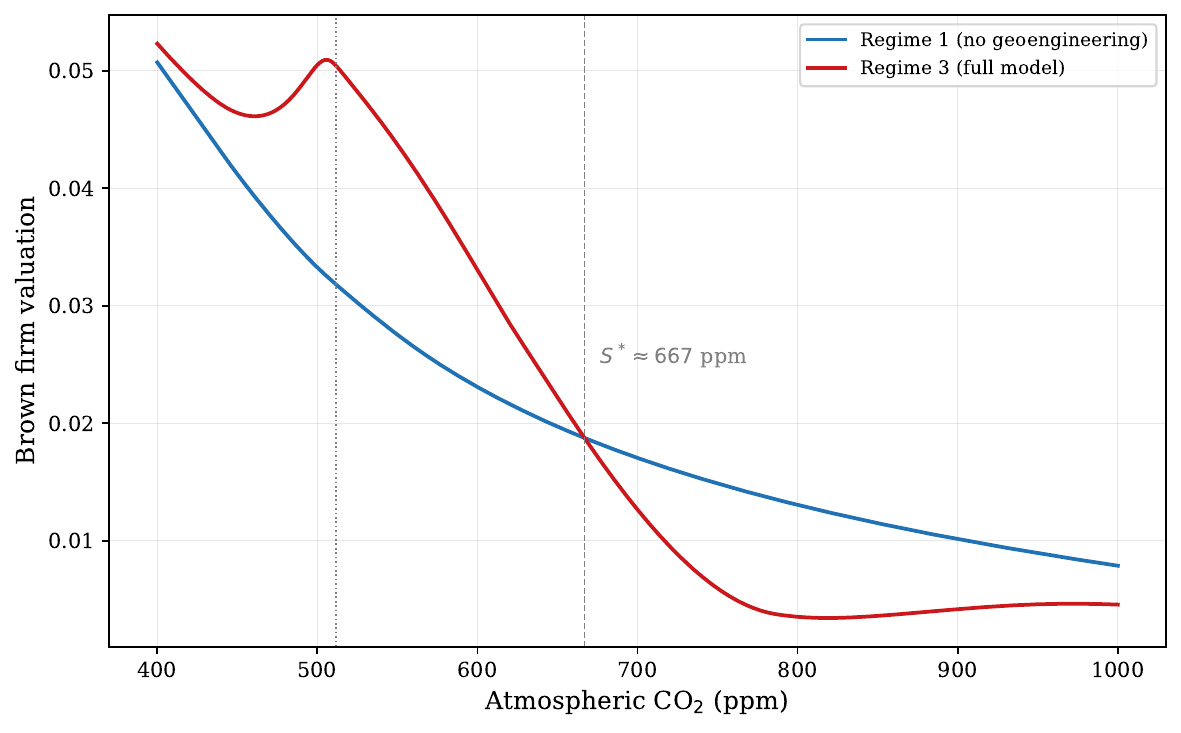}
    \caption{\textbf{Brown Firm Valuations and the Toxic Put.} The vertical axis shows firm value per unit of productivity $A$, with one unit corresponding to the pre-termination annual output benchmark of \$100 trillion. Thus, a value of 0.1 corresponds to \$10 trillion. Brown valuations are shown under Regime 1 and Regime 3, conditional on neither breakthrough nor termination having occurred, while accounting for future events. In our baseline calibration, the Regime 3 curve crosses below the Regime 1 curve at $S^* \approx \InnSstar$ ppm. Below the crossing, geoengineering raises Brown value. Above it, Brown firms are worth less than in a world with no geoengineering at the same carbon stock.}
    \label{fig:brown}
\end{figure}

Figure~\ref{fig:brown} compares Brown firm valuations in Regimes 1 and 3. In Regime 1, valuations fall as the carbon stock grows because the rising carbon tax reduces dividends. In Regime 3, valuations initially rise as deployment approaches and lower carbon taxes become imminent. As ecological debt accumulates, this benefit reverses. The curves cross at $S^*\approx\InnSstar$ ppm, above which Brown firms are worth less with geoengineering than without it. The put has become toxic.

\begin{figure}[htbp]
    \centering
    \includegraphics[width=\textwidth]{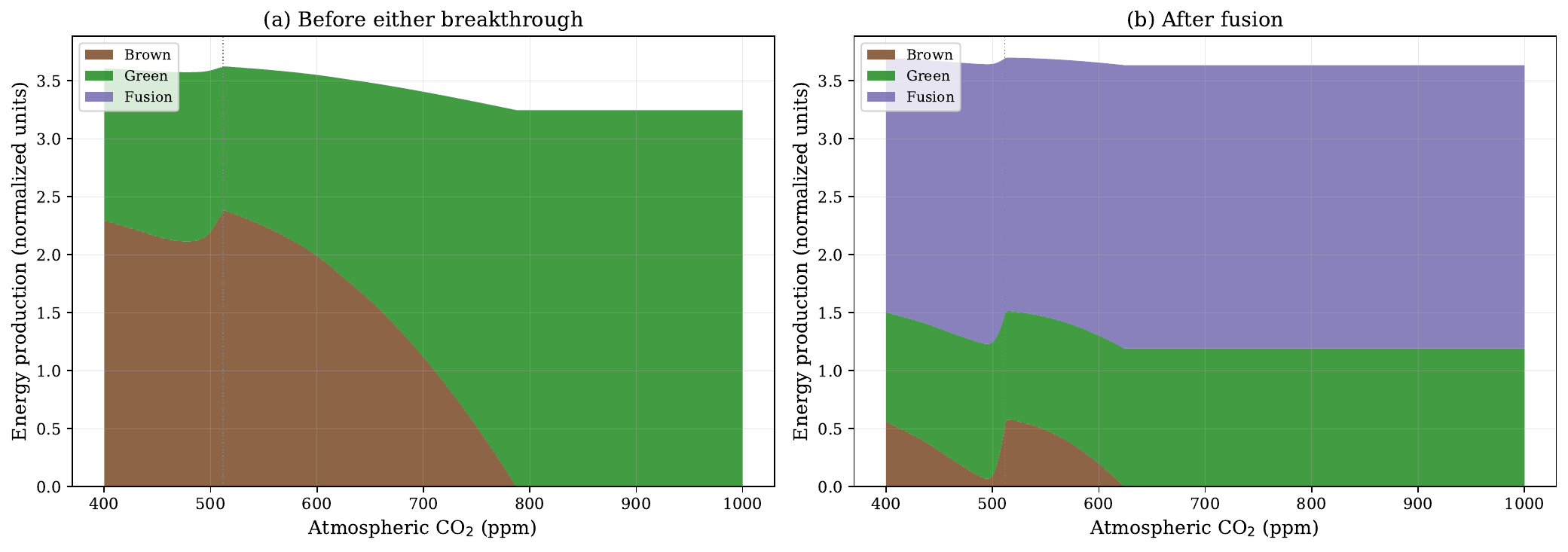}
    \caption{\textbf{Energy Production in Regime 3.} Brown, Green, and fusion production before either breakthrough and after fusion, with CDR unavailable and termination not yet realized. Quantities are in the model's energy units.}
    \label{fig:energy}
\end{figure}

Figure~\ref{fig:energy} connects these valuations to energy production in Regime 3. Before fusion, Brown production falls as the carbon price rises, eventually reaching zero even though Brown assets retain value from possible future price changes. Fusion supplies cheap energy and displaces both Brown and existing Green production. Total energy consumption rises as its price falls.

\subsection{Innovation incentives across regimes} \label{subsec:fig_innovation}

\begin{figure}[htbp]
    \centering
    \includegraphics[width=\textwidth]{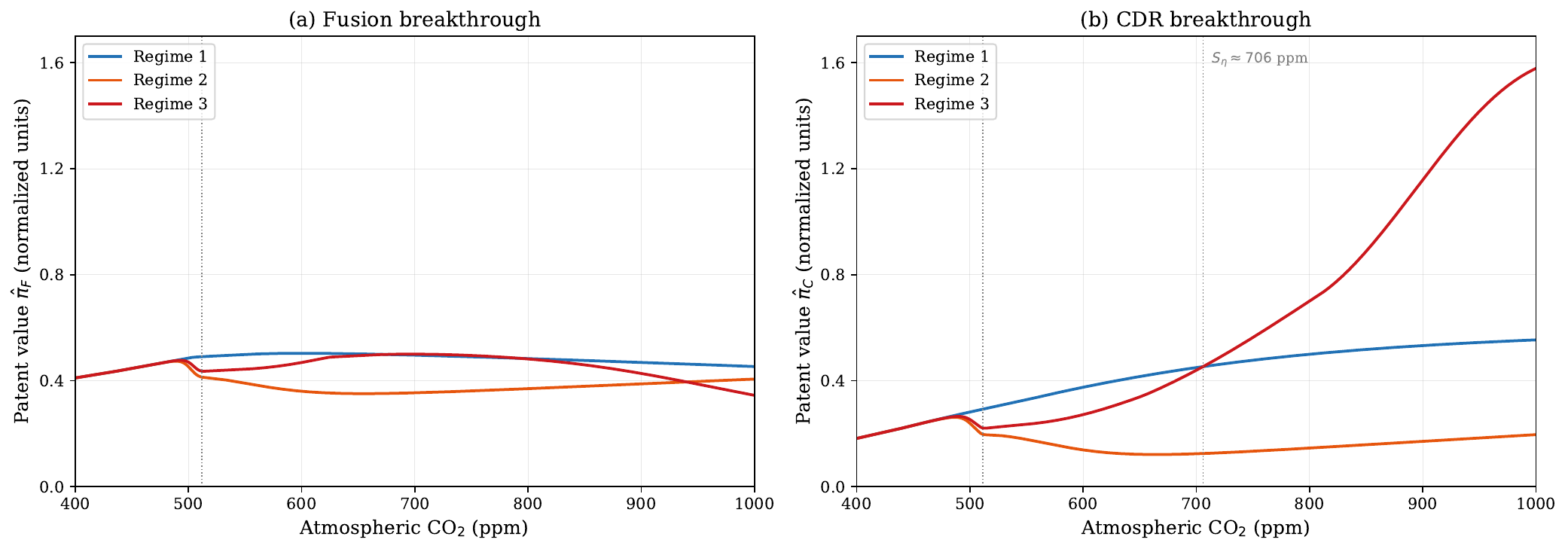}
    \caption{\textbf{Innovation incentives across regimes.} Patent values per unit of productivity, conditional on neither breakthrough nor termination having occurred and accounting for future events. (a) The fusion breakthrough. (b) The CDR breakthrough, with the crossing $S_\eta \approx \InnSeta$ ppm marked with a vertical dotted line.}
    \label{fig:innovation_hazards}
\end{figure}

Figure~\ref{fig:innovation_hazards} displays innovation incentives, measured by patent values $\hat\pi_F(S)$ and $\hat\pi_C(S)$, under the three regimes. The figure quantifies how geoengineering and termination risk affect the innovation rewards described in Section~\ref{subsec:innovation_results}.

Panel (a) shows fusion innovation incentives. In Regime 2, geoengineering lowers carbon taxes, making Brown energy cheaper to supply and reducing energy prices. This reduction in energy prices lowers the profits a new fusion supplier can earn, weakening innovation incentives relative to Regime 1. Termination risk partly offsets this effect in Regime 3. At very high stocks, however, the permanent productivity loss weakens the fusion patent's value. Beyond \GreenRevivalReversal\ ppm, fusion incentives fall below their Regime 2 level.

Panel (b) shows CDR innovation incentives. In Regime 2, geoengineering suppresses the carbon price and the removal reward. In Regime 3, the reward grows with ecological debt. Removal incentives overtake their Regime 1 level at $S_\eta\approx\InnSeta$ ppm, above the toxic put threshold $S^*\approx\InnSstar$ ppm. This means that, throughout the deployment phase in which geoengineering subsidizes fossil fuel firms, geoengineering also depresses removal incentives below their no-geoengineering level.

Comparing the fusion and CDR panels shows how termination risk redirects innovation toward carbon removal. At 650 ppm, moving from Regime 2 to Regime 3 raises the CDR patent value by \CdrPrizeRevival\%, compared with \FusionPrizeRevival\% for fusion. The revival in incentives is nonetheless late. From current concentrations, the mean waiting time for CDR is \ArrCROneFromSzero\ years without geoengineering and \ArrCRThreeFromSzero\ years with it, a delay of approximately \ArrCdrDelayYears\ year. All patent values shown are conditional on neither breakthrough having arrived and no termination yet. Once fusion arrives, its direct boost to CDR discovery also operates.

Table~\ref{tab:signs} summarizes our results so far regarding carbon prices, firm valuations, and innovation incentives under our baseline calibration, before either breakthrough or termination. Without termination risk, geoengineering lowers carbon prices, raises Brown firm values, and weakens innovation incentives. Adding termination risk partly reverses these effects and disproportionately strengthens CDR incentives at moderate carbon stocks. Relative to a world without geoengineering, the effects depend on the accumulated carbon stock. At high stocks, carbon prices and CDR incentives are higher, Brown firms are worth less, and Green firms are worth more. Fusion incentives remain below their no-geoengineering level.

\begin{table}[htbp]
\centering
\caption{Carbon Prices, Firm Values, and Innovation Incentives Across Regimes}
\label{tab:signs}
\begin{tabular}{lccccc}\toprule Comparison & $\tau$ & $V^B$ & $V^G$ & $\hat\pi_F$ & $\hat\pi_C$ \\ \midrule
Regime 2 vs. 1 & $\downarrow$ & $\uparrow$ & $\downarrow$ & $\downarrow$ & $\downarrow$ \\
Regime 3 vs. 2 & $\uparrow$ & $\downarrow$ & $\uparrow$ & $\uparrow^*$ & $\uparrow\uparrow$ \\
Regime 3 vs. 1, low $S$ & $\downarrow$ & $\uparrow$ & $\downarrow$ & $\downarrow$ & $\downarrow$ \\
Regime 3 vs. 1, mod. $S$ & $\downarrow$ & $\uparrow$ & $\downarrow$ & $\downarrow$ & $\downarrow$ \\
Regime 3 vs. 1, high $S$ & $\uparrow$ & $\downarrow$ & $\uparrow$ & $\downarrow$ & $\uparrow$ \\
\bottomrule\end{tabular}

\begin{minipage}{0.94\textwidth}
\vspace{4pt}
{\footnotesize \textit{Notes:} Comparisons use the baseline calibration before either breakthrough or termination. Low, moderate, and high stocks are represented by 550, 650, and 800 ppm. The first two rows summarize comparisons over 400--1,000 ppm. $^*$Fusion incentives are lower in Regime 3 than Regime 2 above \GreenRevivalReversal\ ppm. The double arrow indicates the larger proportional increase in CDR incentives at moderate stocks.}
\end{minipage}
\end{table}

\subsection{Safe Landing and Welfare} \label{subsec:fig_landing} \label{subsec:fig_two_clocks} \label{subsec:fig_frontier} \label{subsec:fig_procurement}

\begin{figure}[htbp]
    \centering
    \includegraphics[width=\textwidth]{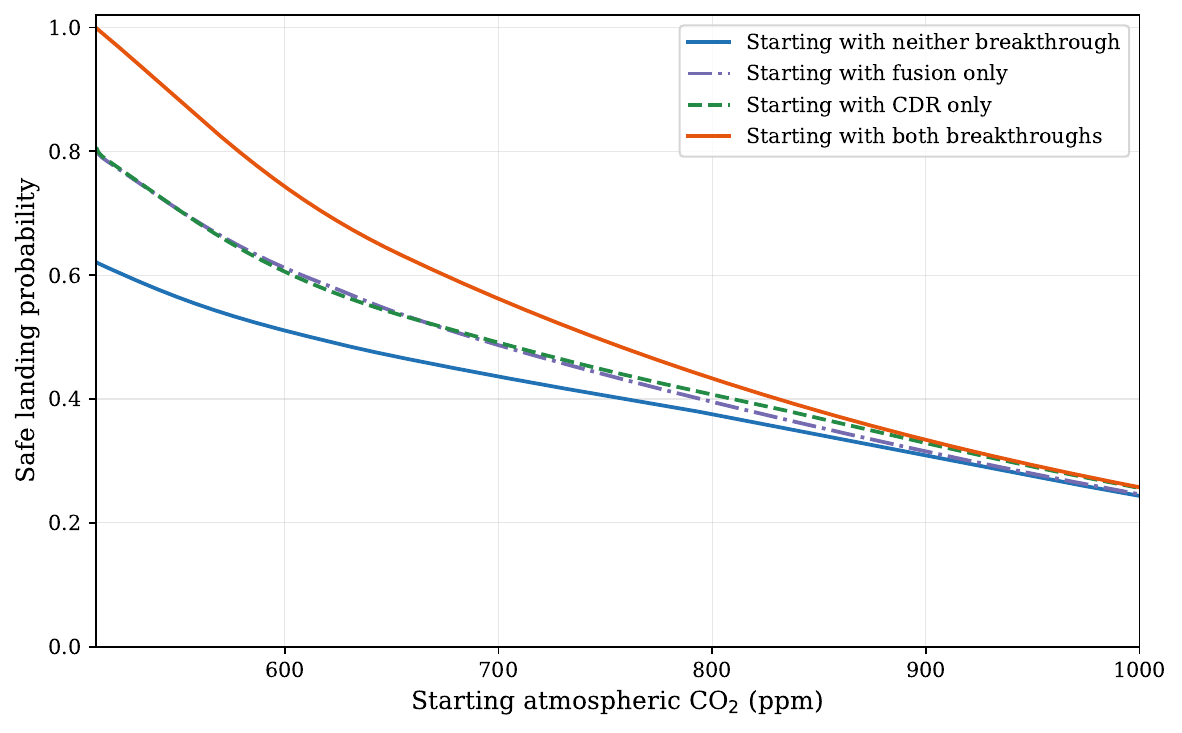}
    \caption{\textbf{The Safe Landing Probability.} Probabilities of shutting down geoengineering with nonpositive net emissions before termination, conditional on prior deployment and no termination yet. The four curves start with neither breakthrough, fusion only, CDR only, or both technologies available. Missing technologies can arrive subsequently.}
    \label{fig:landing}
\end{figure}

Figure~\ref{fig:landing} reports the safe landing probability at each carbon stock for four starting states: neither breakthrough, fusion only, CDR only, or both breakthroughs. Appendix~\ref{app:proofs} describes how we calculate these probabilities. All curves condition on deployment having occurred and geoengineering still operating. A safe landing means returning to the deployment threshold with nonpositive net emissions before termination. With neither technology available, the probability is about 62\% just above the deployment threshold and falls to \LandingNeitherAtSixFiveZero\% at 650 ppm.

With CDR alone or both technologies already available, the probabilities at 650 ppm are \LandingCdrAtSixFiveZero\% and \LandingBothAtSixFiveZero\%, respectively. The fusion-only and CDR-only curves nearly overlap because each technology alone leaves the world waiting for the other: fusion reduces new emissions but cannot remove accumulated carbon, while continued emissions largely offset CDR near the deployment threshold. The expected waiting times for the other missing technology once one is available are also similar in our calibration, partly because fusion makes a CDR breakthrough more likely. The figure shows that discovery of both technologies improves the odds but does not ensure a safe landing: removing accumulated carbon takes decades, and termination can occur during that time.

Appendix Figure~\ref{fig:procurement} shows why inexpensive removal alone may not bring the carbon stock back to the deployment threshold. Removal operates at its annual capacity limit, but new emissions offset it near \StallDownward\ ppm when CDR is available without fusion. Below this balance point, a lower carbon price raises Brown production until emissions exceed removal. Absent a fusion breakthrough or termination, the stock tends to fluctuate around that level.

We now turn to welfare. We compare welfare in Regime 3, where geoengineering is available but subject to termination risk, with welfare in Regime 1, where geoengineering does not exist. This comparison informs the debate over a geoengineering moratorium \citep{Biermann2022}: would the world be better off preventing deployment altogether? Because deployment is triggered by a vulnerable region's threat point rather than chosen by the social planner in our model, the availability of geoengineering need not raise welfare.

\begin{samepage}
We can decompose the welfare difference into the value of buying time and the change associated with innovation incentives:
\begin{equation} \label{eq:decomposition}
    \underbrace{v^{R3}(\eta^{R3}) - v^{R1}}_{\text{net effect}} = \underbrace{\left[v^{R3}(\eta^{R1}) - v^{R1}\right]}_{\text{value of buying time}} + \underbrace{\left[v^{R3}(\eta^{R3}) - v^{R3}(\eta^{R1})\right]}_{\text{deterrence effect}}
\end{equation}
\end{samepage}
In this decomposition, $v^{R3}(\eta^{R1})$ is welfare in an auxiliary Regime 3 economy with breakthrough arrival rates held at their Regime 1 values. We first solve Regime 1 to obtain these arrival-rate functions, then solve Regime 3 again using them at each carbon stock and technology state, including after the first breakthrough. The rates still vary as the carbon stock and available technologies change. Carbon policy and the resulting carbon dynamics are solved again under geoengineering.

The value of buying time is the welfare gain in this auxiliary economy relative to Regime 1, including both the benefits of cooling and the costs of termination risk. The deterrence effect captures the welfare change when innovation incentives respond to geoengineering: we compare the full Regime 3 economy with the auxiliary economy. Either term can be negative. Although we call the second term the deterrence effect, it can be positive when the innovation response raises welfare.\footnote{The decomposition is additive in value-function units. Consumption equivalents satisfy $1+CE_{total}=(1+CE_{buy})(1+CE_{det})$.}

\begin{figure}[htbp]
    \centering
    \includegraphics[width=\textwidth]{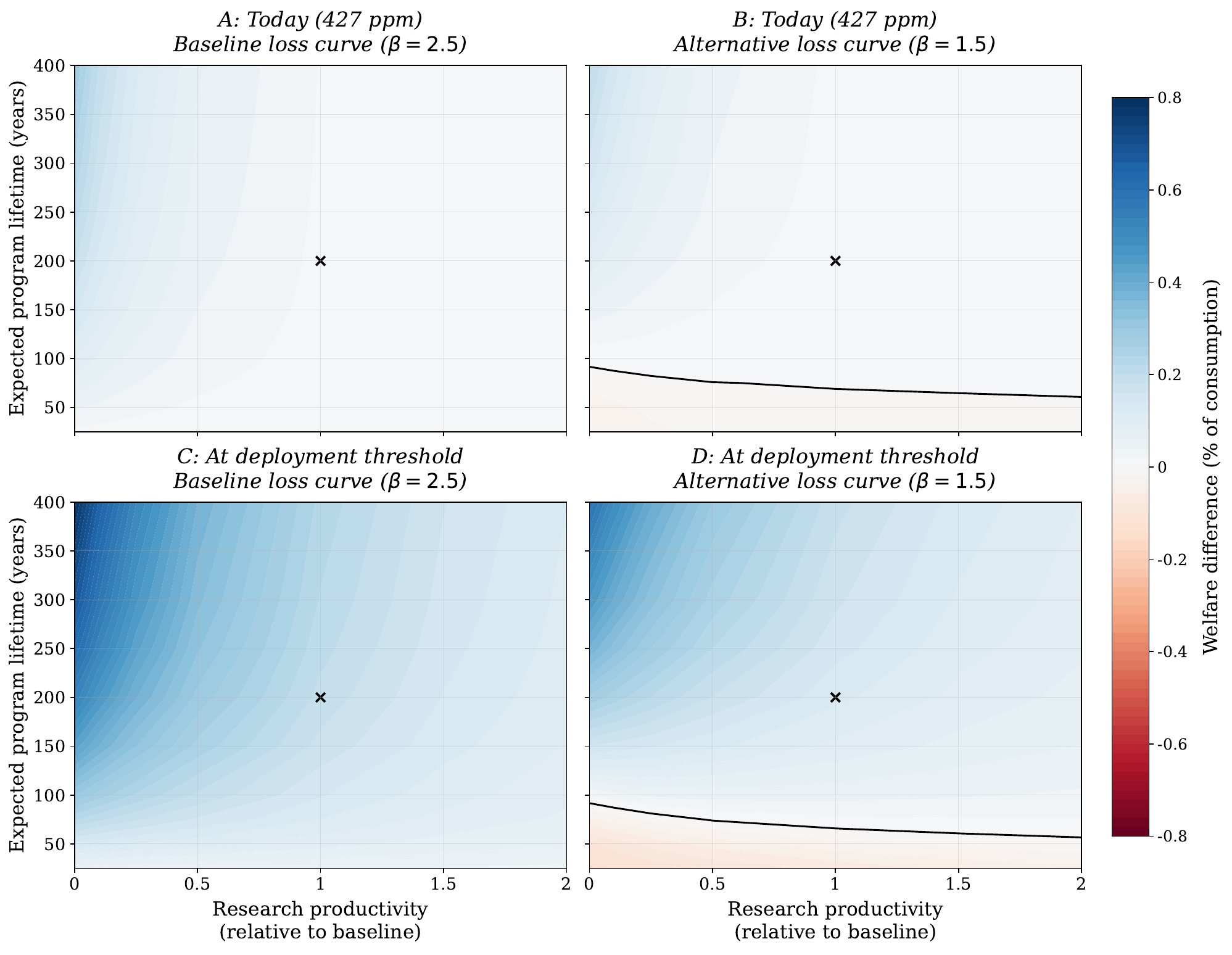}
    \caption{\textbf{Welfare With and Without Geoengineering.} The welfare difference between Regime 3 and Regime 1 in consumption-equivalent percent, conditional on no breakthrough or termination yet. Panels A and B evaluate welfare today (427 ppm), and Panels C and D at the deployment threshold (approximately 512 ppm). Panels A and C use the baseline loss curve ($\beta=2.5$), and Panels B and D use an alternative curve ($\beta=1.5$). Both imply a 12.5\% loss at 800 ppm. The horizontal axis scales both research coefficients relative to baseline. The vertical axis is the inverse termination hazard while cooling operates. Crosses mark baseline research productivity and program lifetime, and the black contours denote zero welfare difference. All panels share the same color scale. Each cell uses 2,243 carbon-grid points.}
    \label{fig:frontier}
\end{figure}

Figure~\ref{fig:frontier} shows the welfare difference between Regime 3 and Regime 1, conditional on no breakthrough or termination yet. The horizontal axis varies research productivity for both technologies by multiplying their baseline research coefficients by the same factor. A value of one gives baseline productivity, while zero removes the response to innovation rewards (although breakthroughs can still occur because the background arrival rates and the additional CDR arrival rate after fusion remain unchanged). The vertical axis varies the expected lifetime of a geoengineering program. A longer lifetime means a lower annual probability of termination. 

Panel A evaluates welfare from the perspective of the present, with a carbon stock of 427 ppm, using the baseline loss curve. We find that welfare is higher with geoengineering throughout the parameter range shown, with the largest gain in the upper left corner. There, breakthroughs arrive more slowly, making the world more reliant on geoengineering to avoid climate damages, while a lower termination probability allows those benefits to last longer.

Panel B lowers the termination-loss exponent from $\beta=2.5$ to $1.5$. This change in the curvature of the termination loss function increases losses at moderate carbon stocks while holding the loss at 800 ppm fixed at 12.5\%. We find that welfare can then be lower with geoengineering when programs are short-lived: termination is more likely, and the productivity loss if termination occurs is larger even before much carbon has accumulated. At baseline research productivity and an expected program lifetime of 25 years, the welfare difference is $-0.006\%$ of consumption.

Panels C and D repeat these welfare comparisons at the deployment threshold of approximately 512 ppm (2$^\circ$C), conditional on neither breakthrough having arrived yet. Welfare differences, both positive and negative, are much larger once the world reaches this threshold. At baseline research productivity and a 200-year expected program lifetime, the gain under the baseline loss curve rises from 0.015\% today to 0.193\% at the deployment threshold. With the alternative loss curve and a 25-year expected program lifetime, the loss grows from 0.006\% to 0.073\%. These differences are larger because, from today's perspective, breakthroughs may arrive before deployment becomes necessary, and the benefits and costs of deployment lie further in the future. At the deployment threshold, cooling begins before either breakthrough has arrived, bringing both the benefits of avoided warming and exposure to future termination losses closer.

\begin{figure}[htbp]
    \centering
    \includegraphics[width=\textwidth]{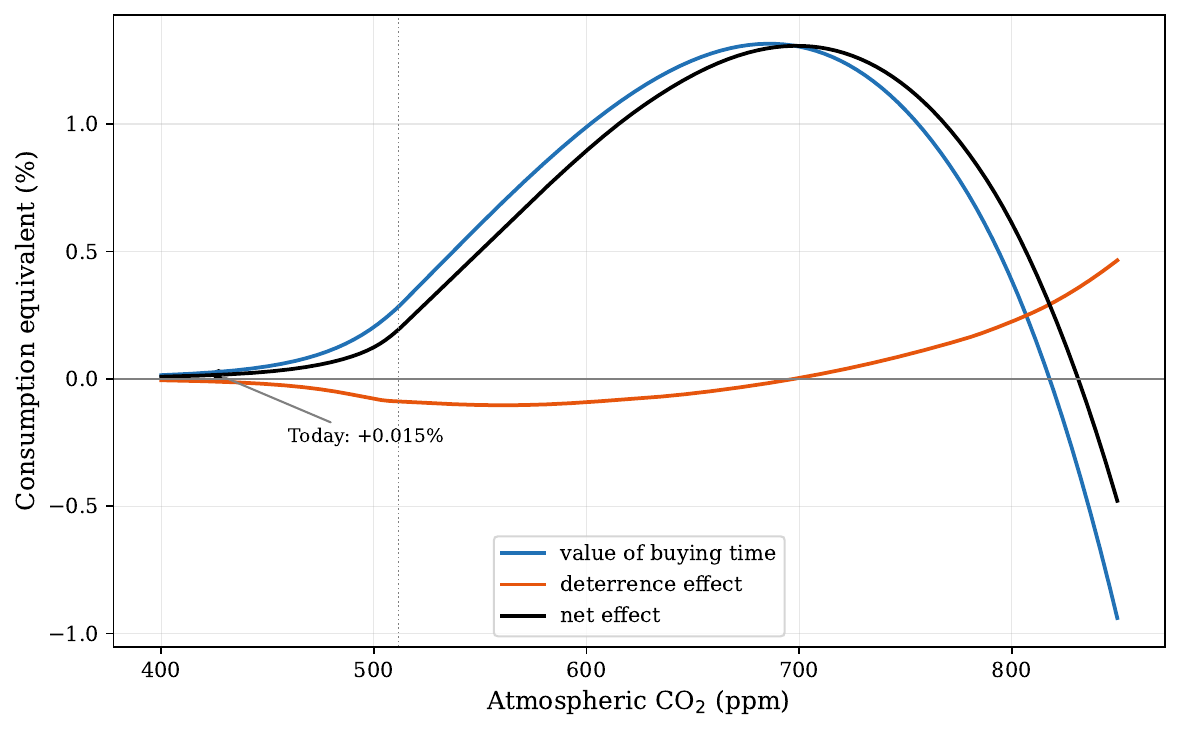}
    \caption{\textbf{The Welfare Decomposition.} Regime 3 continuation welfare relative to Regime 1 at the same stock, before either breakthrough or termination. Above activation, Regime 3 is already deployed.}
    \label{fig:two_clocks}
\end{figure}

Figure~\ref{fig:two_clocks} compares continuation welfare at each carbon stock. Above the deployment threshold, the Regime 3 economy has already deployed geoengineering. The welfare difference is decomposed into the value of buying time and the deterrence effect, as in Equation~(\ref{eq:decomposition}). The value of buying time peaks around \WelBridgePeakStock\ ppm, then declines as ecological debt grows. The deterrence effect becomes positive above \WelDeterrencePositiveAbove\ ppm, meaning that the overall change in innovation incentives from geoengineering raises welfare. Termination risk's revival of CDR innovation incentives contributes to this gain. Despite that improvement, the total welfare difference becomes negative above \WelNetNegativeAbove\ ppm.

We next assess welfare under alternative deployment rules. Our baseline takes the deployment function $g(S)$ as exogenous, motivated by a vulnerable nation's credible threat of unilateral deployment. An alternative would be to let the social planner choose when and how much to deploy. We therefore consider earlier deployment and partial offsets of further warming, holding other parameters at baseline values. Appendix Figure~\ref{fig:deployment_rules} illustrates these alternative deployment rules. Each offsets a fixed share of further warming after deployment and prevents temperatures from exceeding 2$^\circ$C (so the vulnerable nation's threat still binds).

\begin{figure}[htbp]
\centering
\includegraphics[width=\textwidth]{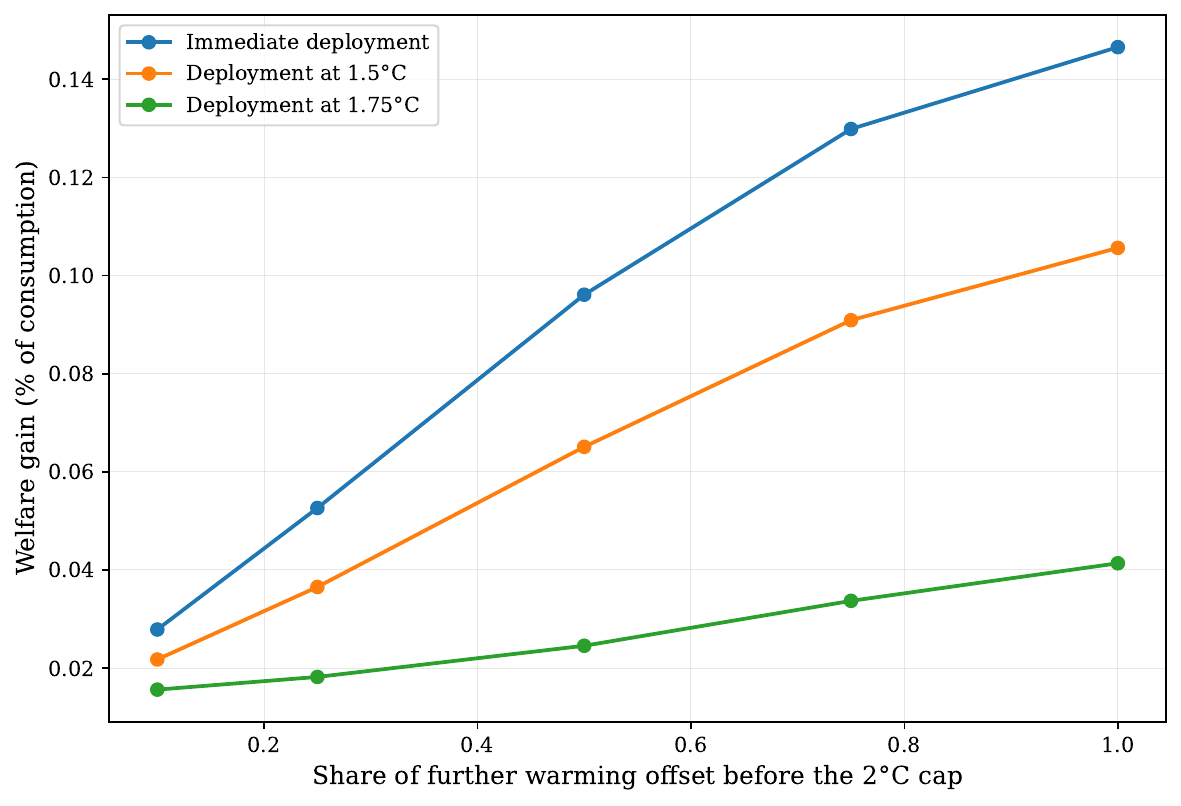}
\caption{\textbf{Welfare Under Alternative Deployment Rules.} Welfare gains relative to Regime 1, evaluated today, with all other parameters at baseline values. Each rule offsets a fixed share of further warming after deployment and prevents temperatures from exceeding 2$^\circ$C. A share of one prevents all further warming. Among the rules considered, earlier deployment and a larger share of warming offset both raise welfare.}
\label{fig:partial_deployment_welfare}
\end{figure}

Figure~\ref{fig:partial_deployment_welfare} compares welfare across these rules. At each deployment threshold considered, welfare is highest when geoengineering prevents all further warming. Earlier deployment also raises welfare. Immediate deployment that prevents all further warming yields a gain of about 0.15\% of consumption relative to Regime 1, compared with 0.015\% under the baseline rule. Stabilizing at 1.5$^\circ$C raises welfare by about 0.11\% of consumption, lowers the carbon tax by about 29\%, and raises Brown firm value by about 30\%.

These results imply that, among the rules we consider, the social planner would choose to deploy geoengineering immediately and prevent all further warming.\footnote{Our model does not allow partial deployment to generate knowledge that improves subsequent deployment or reduces its risks. In practice, a smaller pilot program could help the world learn how to operate geoengineering more effectively and safely. These benefits could favor a more gradual deployment path.} Yet the world is not deploying geoengineering today. This gap suggests that frictions outside our model may prevent deployment from following the social planner's preferred rule. We therefore focus on a potentially more realistic trigger: a vulnerable nation's credible threat of unilateral deployment.

\subsection{Risk Premia Across Firms} \label{subsec:fig_trp_decomp} \label{subsec:cross_section}

\begin{figure}[htbp]
    \centering
    \includegraphics[width=\textwidth]{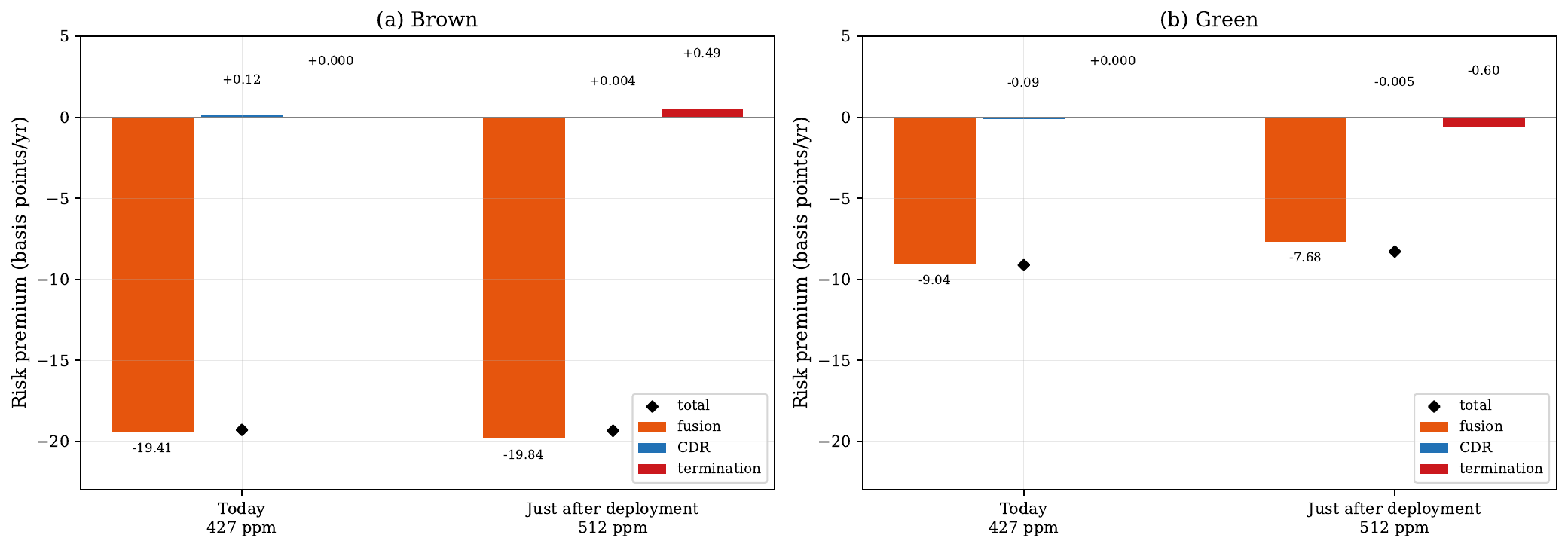}
    \caption{\textbf{Risk Premia Today and at Deployment.} Jump-risk components in Regime 3 at 427 and 512 ppm, before either breakthrough or termination. Values are in basis points per year. Diamonds show the total.}
    \label{fig:trp_decomposition}
\end{figure}

Figure~\ref{fig:trp_decomposition} decomposes the expected excess returns of Brown and Green firms into the three jump components of Equation~(\ref{eq:expected_return}): termination, the fusion breakthrough, and the CDR breakthrough. All values are computed in the Regime 3 economy in which none of these events has yet occurred, at current concentrations and just after deployment, in basis points per year. The sign of each risk premium component depends on how the asset's value moves with consumption at each event. An asset that gains value at an event that lowers consumption is a hedge and earns a negative premium. An asset that gains value at an event that raises consumption earns a positive premium.

We focus on risk premia at current concentrations and at the onset of deployment, when Brown firms still have substantial value. For Brown firms, the termination component is zero today because deployment has not begun, so termination risk is zero. Just after deployment, it is approximately $\BrownTerminationPremiumDeployment$ basis points per year: Brown firms lose value when termination lowers consumption. Green firms also have a zero termination component today but provide a termination hedge just after deployment, gaining value when consumption falls and earning a premium of about $-0.6$ basis points. These termination premia are limited near deployment, because ecological debt is low, although both valuations already account for possible future termination.

Fusion news raises consumption while displacing Brown and existing Green production, giving both assets negative fusion-risk premia. For Brown firms, the fusion component is approximately $\BrownFusionPremiumToday$ basis points per year today and $\BrownFusionPremiumDeployment$ just after deployment. For Green firms, the corresponding values are about $-9.0$ and $-7.7$ basis points. These magnitudes are much larger than the termination components at current concentrations and just after deployment. The fusion arrival rate is higher than the termination hazard. In addition, fusion immediately raises consumption and sharply reduces both firms' values. By contrast, termination is impossible before deployment and causes a relatively limited consumption loss just afterward, when little ecological debt has accumulated. At higher carbon stocks, termination can cause much larger consumption losses.

The CDR risk premium component is much smaller but is not zero. Today, it is about $+0.12$ basis points for Brown firms and $-0.09$ for Green firms; just after deployment, both are below 0.01 basis points in absolute value. A CDR breakthrough lowers the carbon price, raising Brown value and lowering Green value at these stocks. But unlike fusion, it does not immediately change energy technology or remove the accumulated carbon stock. Removal takes time. With zero removal costs, the immediate consumption change comes from adjustments in carbon taxation and energy production, and is much smaller than after fusion. The resulting risk premium can therefore be close to zero even when CDR discovery meaningfully changes firm values.

Appendix Figure~\ref{fig:trp_stocks} shows how these risk premia components vary over the full carbon-stock range. At moderate carbon stocks, Green firms provide disaster insurance through a negative termination component. The insurance sign reverses near \GreenTermTRPCross\ ppm, where productivity destruction and the shrinking gap between pre-termination and post-termination carbon taxes turn the termination component positive. For Brown firms, the termination component initially rises as ecological debt accumulates, but becomes negative at sufficiently high stocks. There, termination can lower the carbon tax enough to restart Brown production, raising equity value even as consumption falls. The CDR components also become larger at higher carbon stocks.

\subsection{Heterogeneity} \label{subsec:robustness}

Table~\ref{tab:robustness} reports welfare, carbon-tax, and Brown-value effects at current concentrations and at the deployment threshold under alternative assumptions. Each row changes the stated assumption in both regimes. Other parameters remain at baseline values.

\begin{table}[htbp]
\centering
\caption{Heterogeneity: Regime 3 vs. Regime 1}
\label{tab:robustness}
\footnotesize
\setlength{\tabcolsep}{4pt}
\begin{tabular}{@{}p{0.36\textwidth}rrrrrr@{}}\toprule
 & \multicolumn{3}{c}{Today (427 ppm)} & \multicolumn{3}{c}{At deployment threshold} \\
 & \multicolumn{3}{c}{} & \multicolumn{3}{c}{($\approx512$ ppm)} \\
\cmidrule(lr){2-4}\cmidrule(l){5-7}
Assumption & Welfare & \shortstack{Carbon\\tax} & \shortstack{Brown\\value} & Welfare & \shortstack{Carbon\\tax} & \shortstack{Brown\\value} \\
 & (CE \%) & (\%) & (\%) & (CE \%) & (\%) & (\%) \\ \midrule
\raggedright Baseline & +0.0151 & -3.37 & +6.54 & +0.1929 & -45.14 & +58.51 \\
\raggedright Carbon tax: 75\% of SCC & +0.0201 & -3.99 & +6.75 & +0.2171 & -44.91 & +41.88 \\
\raggedright Carbon tax: 50\% of SCC & +0.0379 & -6.30 & +7.31 & +0.2958 & -44.88 & +27.29 \\
\raggedright Climate-damage coefficients: $1.5\times$ baseline & +0.0098 & -1.91 & +5.47 & +0.2392 & -41.31 & +103.51 \\
\raggedright Higher background breakthrough rates: 1\%/year each & +0.0080 & -2.25 & +4.09 & +0.1684 & -43.81 & +54.96 \\
\raggedright Constant breakthrough rates: 40/50-year waits & +0.0221 & -5.04 & +8.82 & +0.2899 & -46.35 & +63.99 \\
\raggedright No fusion or CDR breakthroughs & +0.2209 & -13.67 & +37.25 & +0.5590 & -32.73 & +74.27 \\
\raggedright Lower CDR capacity: 10 GtCO$_2$/year & +0.0186 & -3.94 & +9.31 & +0.2298 & -43.97 & +71.10 \\
\raggedright No permanent productivity loss at termination & +0.0185 & -4.11 & +8.30 & +0.2280 & -51.17 & +80.59 \\
\raggedright Termination productivity loss at 800 ppm: 25\% & +0.0124 & -2.76 & +5.18 & +0.1632 & -39.88 & +42.75 \\
\raggedright Increasing termination hazard: $(0.005+0.005g)$/year & +0.0132 & -2.94 & +5.57 & +0.1716 & -41.41 & +47.19 \\
\raggedright Decreasing termination hazard: $(0.001+0.004e^{-g/(1^\circ\mathrm C)})$/year & +0.0170 & -3.78 & +7.49 & +0.2127 & -48.53 & +70.21 \\
\raggedright Higher background breakthrough rates: 0.5\%/year each & +0.0120 & -2.93 & +5.50 & +0.1838 & -44.63 & +57.15 \\
\raggedright Extra background CDR rate after fusion: 0.5\%/year & +0.0148 & -3.30 & +6.57 & +0.1946 & -45.79 & +59.48 \\
\raggedright No extra background CDR rate after fusion & +0.0143 & -3.20 & +6.59 & +0.1971 & -46.48 & +60.49 \\
\raggedright Termination loss exponent: $\beta=1.5$ & +0.0096 & -2.13 & +3.88 & +0.1255 & -31.36 & +32.03 \\
\raggedright Asset discount rate: 3\% (baseline 4\%) & +0.0091 & -2.41 & +5.19 & +0.1675 & -44.62 & +53.60 \\
\bottomrule\end{tabular}

\begin{minipage}{\textwidth}
\vspace{4pt}
{\footnotesize \textit{Notes:} Welfare is Regime 3 relative to Regime 1 in consumption-equivalent percent. Tax and Brown value are percentage differences. The first three columns evaluate effects at 427 ppm and the last three at the deployment threshold (approximately 512 ppm), before either breakthrough or termination. Each comparison uses the same carbon stock in both regimes. All rows use the baseline grid of 4,485 points. Partial-tax cases set both the emissions tax and removal credit to the stated fraction of the SCC in each regime, so the percentage tax effect also equals the percentage SCC effect. The stronger-damage case multiplies $\omega_2$, $\zeta$, and $\omega_{acid}$ by 1.5. Higher background rates increase $a_F$ and $a_C$ from 0.2\% to 0.5\% or 1\% per year, leaving reward sensitivity and the additional CDR rate after fusion unchanged. The two post-fusion cases reduce the extra CDR arrival rate from 1 percentage point per year to 0.5 or zero. The no-permanent-loss case sets $\Gamma_0=0$ but retains renewed warming after termination. The constant-rate case sets $\eta_F=1/40$ and $\eta_C=1/50$ per year until discovery, with no response to rewards or post-fusion CDR boost. The no-breakthrough case sets all arrival rates to zero. The $\beta=1.5$ case retains a 12.5\% permanent loss at 800 ppm. Both alternative termination hazards apply while geoengineering operates. The decreasing hazard starts at 0.5\% per year and approaches a positive 0.1\% floor as cooling increases. Baseline CDR capacity is 15 GtCO$_2$/year. The higher-loss case raises the permanent loss at 800 ppm from 12.5\% to 25\%, keeping the loss exponent fixed. The asset-discount-rate case lowers $r$ from 4\% to 3\%, leaving the welfare discount rate $\rho=2\%$ unchanged.}
\end{minipage}
\end{table}

We first consider carbon taxes equal to 75\% or 50\% of the SCC, allowing for political or other frictions that prevent governments from fully pricing the marginal damages from emissions. We also consider stronger climate damages by multiplying all ordinary damage coefficients by 1.5. This is close to the benchmark used by \citet{OECD2025ClimateScenarios}, based on \citet{HowardSterner2017}.

Several variations explore uncertainty about technological progress. Higher background breakthrough rates, at 0.5\% or 1\% per year for both technologies, represent scientific progress that is less sensitive to market rewards. Constant breakthrough rates with expected waiting times of 40 years for fusion and 50 years for CDR remove the response to innovation incentives. The case with no breakthroughs considers a world in which neither technology becomes available. We also reduce or eliminate the additional CDR arrival rate after fusion to allow for a weaker link between cheap energy and carbon removal technology.

Lower CDR capacity allows for tighter limits on how quickly accumulated carbon can be removed. We also vary termination losses because their magnitude is poorly understood: one case removes the permanent productivity loss but retains the damages from renewed warming, while another doubles the productivity loss at 800 ppm to 25\%. Lowering the loss exponent to $\beta=1.5$ instead raises losses at moderate carbon stocks while keeping the loss at 800 ppm fixed. We also allow the termination hazard to rise or fall with cooling. An increasing hazard captures the possibility that deployment at greater scale leads to unforeseen harms that force termination. A decreasing hazard reflects stronger efforts to prevent interruptions as ecological debt grows. Finally, a lower asset discount rate allows us to assess how discounting affects firm values and inventor rewards, while keeping the welfare discount rate unchanged.

Across all the variations in Table~\ref{tab:robustness}, welfare is higher, carbon taxes are lower, and Brown firm value is higher in Regime 3 than in Regime 1, both today and at the deployment threshold. These results support our conclusion that the availability of the geoengineering put can raise welfare under a range of assumptions. This conclusion is not universal, however. As Figure~\ref{fig:frontier} showed earlier, welfare can be lower with geoengineering when a very high termination hazard is combined with substantial termination losses at moderate carbon stocks.

\section{Conclusion} \label{sec:conclusion}

By providing a cheap and fast-acting way to cool the planet without reducing the carbon stock, geoengineering introduces risks that are sharply different from transition and biodiversity risks studied in the existing climate-finance literature. In this paper, we develop a general equilibrium macro-finance framework in which geoengineering acts as a put option on climate damages, funded by a growing ecological debt of masked warming. If the program terminates before that debt has been paid, temperatures snap back, potentially causing a catastrophic termination shock. Our model allows us to evaluate the argument that geoengineering improves welfare by buying time while the world waits for a climate technology breakthrough.

Our analysis yields four main results. First, geoengineering without termination risk suppresses marginal temperature damages. In our calibration, it lowers carbon prices, raises fossil fuel valuations, and reduces the rewards to carbon removal.

Second, termination risk changes the cross-section of asset prices. The prospect of geoengineering initially benefits fossil fuel firms by suppressing the carbon tax, but as ecological debt builds, those same assets become toxic: their value falls below its level in a world without geoengineering. At high carbon stocks, the social cost of carbon can also exceed its level without geoengineering. The initial reprieve from carbon taxation therefore comes with exposure to future termination. Fusion creates a different risk, lowering both Brown and existing Green firm values through competition from cheap, clean energy.

Third, geoengineering and carbon removal can switch from substitutes to complements when termination risk is present, because carbon removal can retire the ecological debt that accumulates under geoengineering. Policymakers should therefore evaluate geoengineering jointly with carbon removal, rather than treating one as a substitute for the other.

Fourth, geoengineering and the associated termination risk direct innovation. Deployment lowers realized temperatures while the world waits for a breakthrough. But it weakens the incentive to innovate, because innovators earn profits tied to energy and carbon prices that deployment can suppress. We consider two breakthroughs the world might wait for. A fusion breakthrough reduces emissions but leaves the accumulated carbon in place. Carbon removal is the only technology that can retire ecological debt, and termination risk disproportionately revives its innovation incentives at moderate carbon stocks. Our analysis also shows that carbon removal has a built-in incentive problem: it is paid through the price of carbon, and that price falls as removal succeeds. Even after discovery, limited annual capacity and new emissions leave the world exposed to termination while removal works down the stock. A breakthrough therefore does not ensure a safe landing, in which the world returns to the deployment threshold with nonpositive net emissions before termination occurs.

We find that welfare is higher with geoengineering in our baseline calibration and across a range of alternative parameter assumptions, both today and at the deployment threshold. Under our baseline calibration, a breakthrough often arrives before deployment starts. When deployment does begin, the capped temperatures buy decades of avoided damages before ecological debt becomes large. The welfare ranking is not universal, however. Welfare can be lower with geoengineering when a very high termination hazard is combined with substantial termination losses at moderate carbon stocks.

More broadly, our model offers a framework for thinking through the consequences of the geoengineering put. Our calibration results depend on parameters that remain highly uncertain, including the termination hazard, the severity of the termination shock, and the timing and capabilities of breakthrough technologies. That uncertainty points directly to the need for further research. Research done today will help determine whether geoengineering should be deployed, and if so, how. As the physical conditions for deployment strengthen, the asset pricing and innovation predictions of the model may become testable in market data. We leave that empirical work for future research.

\newpage

\appendix

\section{The Landing-Probability System} \label{app:proofs}

Let $z$ record technology availability, $\mu_z(S)=B_z(S)-R_z(S)$, and $p_z$ be the probability of landing before termination, conditional on prior deployment. In states outside the landing set,
\begin{equation}\label{eq:landing_pre}
0=\tfrac12\sigma_S^2 p_z''+\mu_z p_z'-\lambda p_z+
\sum_{i\text{ not yet available}}\eta_i(S,z)\,[p_{z+i}-p_z].
\end{equation}
The landing boundary has $p_z=1$ where $g(S)=0$ and $\mu_z(S)\leq0$. Termination is a failure, with value zero, even though innovation and carbon removal can continue afterward. We solve all four technology states on the same grid. Each discovery moves to the corresponding continuation probability rather than counting automatically as a landing. Noise can later push carbon above the deployment threshold, so this is a first-passage probability rather than a claim of permanent safety.

\section{Supplementary Quantitative Results}\label{app:dynamics}

\begin{figure}[htbp]
\centering
\includegraphics[width=\textwidth]{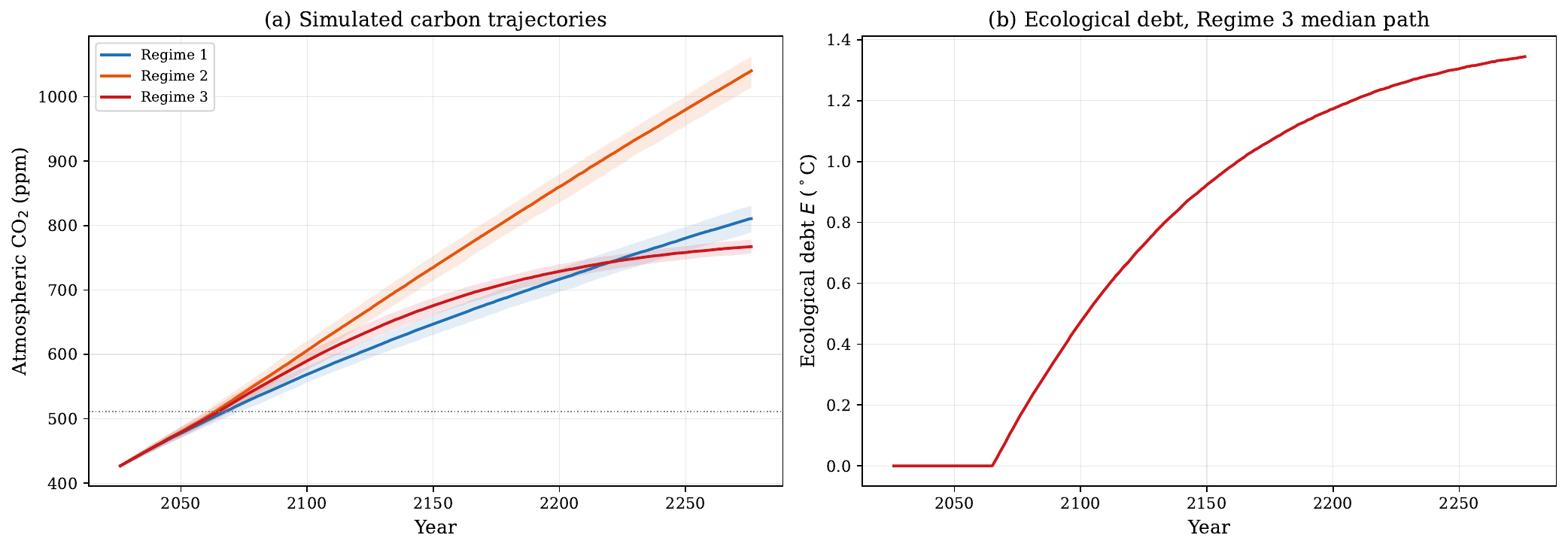}
\caption{\textbf{Illustrative Carbon Dynamics.} Simulated paths with breakthrough and termination events suppressed and full-model pre-event policies retained.}
\label{fig:dynamics}
\end{figure}

Figure~\ref{fig:dynamics} summarizes 600 simulated carbon-stock paths per regime over 250 years, starting from today's stock and using annual time steps. Panel (a) shows the median paths and shaded bands covering the 5th to 95th percentiles. Panel (b) shows ecological debt along the Regime 3 median path. The simulations include random carbon-stock fluctuations but suppress breakthrough and termination events, while retaining policies that account for their future possibility.

\clearpage\subsection{Emissions and Removal Capacity}\label{app:removal}

\begin{figure}[htbp]
    \centering
    \includegraphics[width=\textwidth]{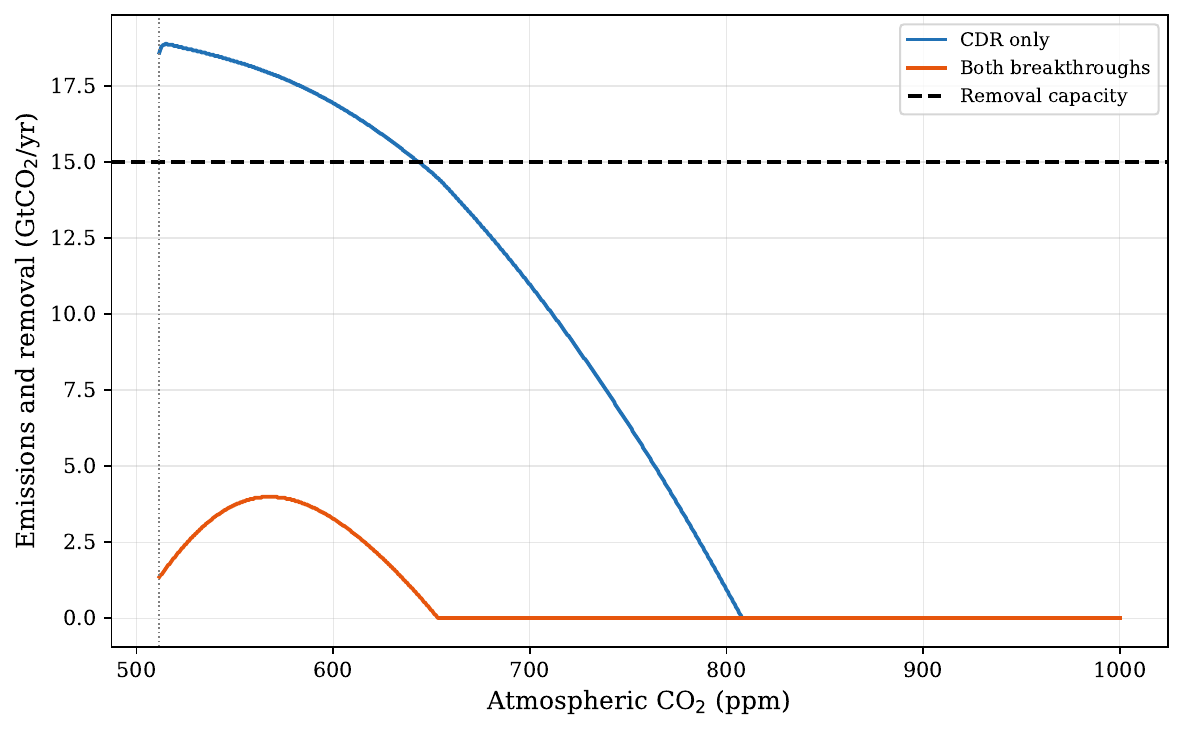}
    \caption{\textbf{Emissions and Removal Capacity.} Regime 3 emissions with CDR alone and with both technologies, compared with annual removal capacity. With CDR alone, emissions equal removal near \StallDownward\ ppm.}
    \label{fig:procurement}
\end{figure}

\clearpage\subsection{Risk Premia Across Carbon Stocks}\label{app:premia}
\begin{figure}[htbp]
\centering
\includegraphics[width=\textwidth]{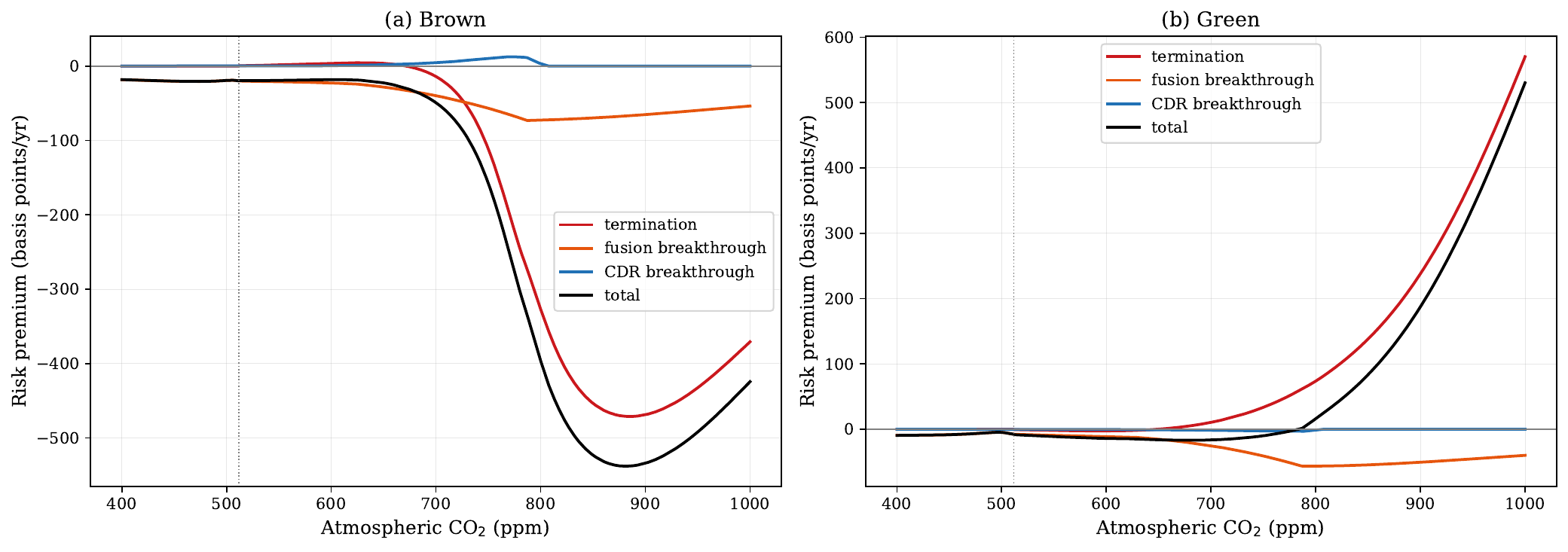}
\caption{\textbf{Risk Premia Across Carbon Stocks.} The same three jump components as Figure~\ref{fig:trp_decomposition}, shown over the full reporting range before either breakthrough or termination.}
\label{fig:trp_stocks}
\end{figure}

\clearpage\subsection{Earlier Deployment Rules}
\begin{figure}[htbp]
\centering
\includegraphics[width=\textwidth]{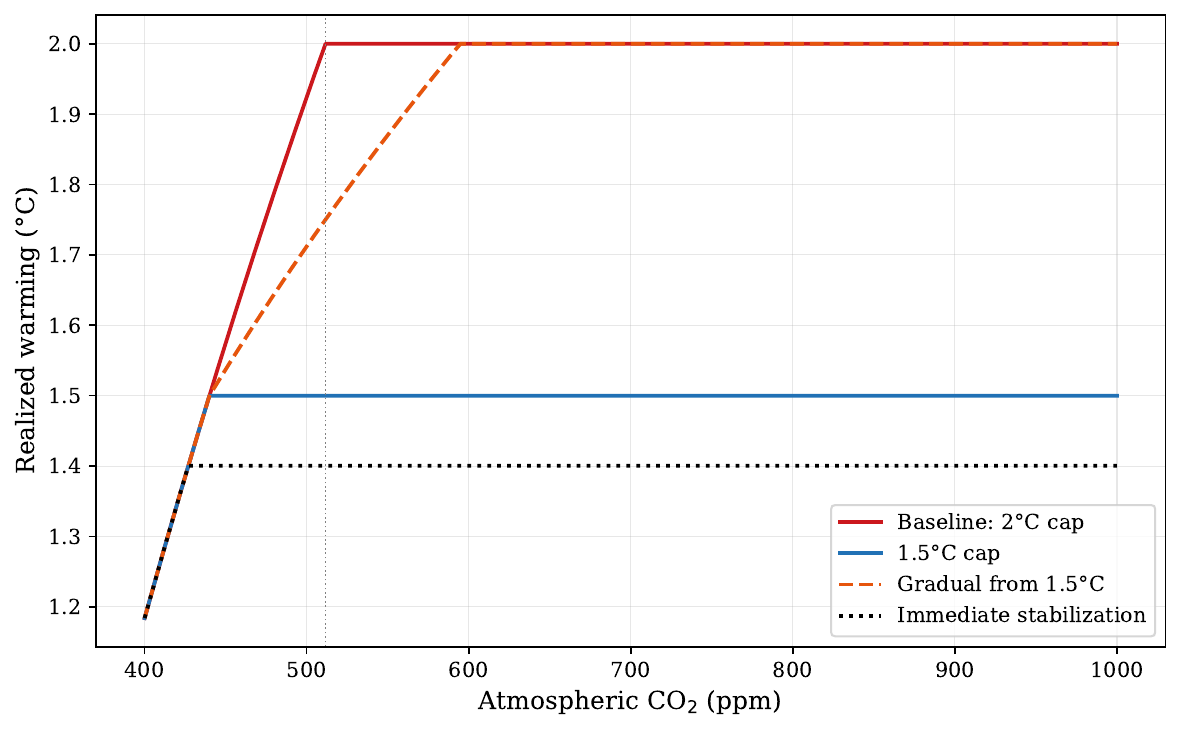}
\caption{\textbf{Alternative Deployment Rules.} Rules begin without an initial temperature reduction and cap warming at or below 2$^\circ$C.}
\label{fig:deployment_rules}
\end{figure}

\end{document}